\documentclass[
  aps,
  prb,
  reprint,
  superscriptaddress,
  amsfonts,amssymb,amsmath
]{revtex4-2}

\usepackage{bm}
\usepackage{latexsym}
\usepackage{physics}
\usepackage[dvipdfmx]{graphicx}
\usepackage{hyperref}
\hypersetup{
  setpagesize  = false,
  colorlinks   = true,
  urlcolor     = blue,
  linkcolor    = blue,
  citecolor    = blue
}

\begin{document}

%%% title %%%
\title{Entanglement swapping in critical quantum spin chains}

%%% author(s) %%%
\author{Masahiro Hoshino}
\email{hoshino-masahiro921@g.ecc.u-tokyo.ac.jp}
\affiliation{Department of Physics, University of Tokyo, 7-3-1 Hongo, Bunkyo-ku, Tokyo 113-0033, Japan}

\author{Masaki Oshikawa}
\affiliation{Institute for Solid State Physics, University of Tokyo, Kashiwa, Chiba 277-8581, Japan}
\affiliation{Kavli Institute for the Physics and Mathematics of the Universe (WPI), University of Tokyo, Kashiwa, Chiba 277-8581, Japan}

\author{Yuto Ashida}
\affiliation{Department of Physics, University of Tokyo, 7-3-1 Hongo, Bunkyo-ku, Tokyo 113-0033, Japan}
\affiliation{Institute for Physics of Intelligence, University of Tokyo, 7-3-1 Hongo, Bunkyo-ku, Tokyo 113-0033, Japan}

%%% date %%% 
\date{\today}

%%% abstract %%%
\begin{abstract}
  The transfer of quantum information between many-qubit states is a subject of fundamental importance in quantum science and technology.
  We consider entanglement swapping in critical quantum spin chains, where the entanglement between the two chains is induced solely by the Bell-state measurements.
  We employ a boundary conformal field theory (CFT) approach and describe the measurements as conformal boundary conditions in the replicated field theory.
  We show that the swapped entanglement exhibits a logarithmic scaling, whose coefficient takes a universal value determined by the scaling dimension of the boundary condition changing operator.
  We apply our framework to the critical spin-$\frac{1}{2}$ XXZ chain and determine the universal coefficient by the boundary CFT analysis.
  We also numerically verify these results by the tensor-network calculations.
  Possible experimental relevance to Rydberg atom arrays is briefly discussed.
\end{abstract}

\maketitle

%%%========== main text  ==========%%%
\section{Introduction}
Quantum information transfer and teleportation are central themes in quantum science and technology~\cite{bennett1993teleporting,gottesman2012longerbaseline}.
One of the key concepts there is \emph{entanglement swapping}, a process that entangles two systems solely by quantum measurements~\cite{zukowski1993eventreadydetectors,bose1998multiparticle,hardy2000entanglementswapping,horodecki2009quantum,fan2009entanglement,song2014purifying,bej2022creating,ji2022entanglement,ji2022entanglementa,zangi2023entanglement,bej2024activation,dmello2024entanglementswapping}.
After two decades from its first realization~\cite{pan1998experimental}, recent advances~\cite{metcalf2014quantum,northup2014quantum,pirandola2015advances,covey2023quantum,hu2023progress} have now enabled quantum information transfer over distances exceeding $1000$ km with high fidelity~\cite{ren2017groundtosatellite,chen2021integrated}, showing great promise for realizing quantum cryptography~\cite{pirandola2020advances} and quantum networks~\cite{cirac1997quantum,briegel1998quantum,kimble2008quantum,wehner2018quantum}.

Transfer of quantum information has also been an intriguing theoretical subject, dating back to the pioneering works on the Bell inequality~\cite{bell1964einstein} and the Lieb-Robinson bound~\cite{lieb1972finite}.
An early study~\cite{bose1998multiparticle} has highlighted the possible advantage of entanglement swapping in multiparticle setups, and subsequent studies have explored the potential of using many-body systems for quantum information transfer~\cite{bose2003quantum,bose2007quantum,bayat2010informationtransferring,sala2024quantum,eckstein2024robust}.
From a broader perspective, previous studies have revealed various aspects of many-body states subject to measurement backaction; examples include measurement-induced phase transitions in quantum circuits~\cite{bao2020theory,chan2019unitaryprojective,gullans2020dynamical,jian2020measurementinduced,li2018quantum,li2019measurementdriven,li2021conformal,skinner2019measurementinduced,zabalo2022operator,fisher2023random} and monitored fermions/bosons~\cite{alberton2021entanglement,cao2019entanglement,chen2020emergent,fava2023nonlinear,fuji2020measurementinduced,Gopalakrishnan2021,Turkeshi2021,jian2022criticality,kawabata2023entanglement,ladewig2022monitored,merritt2023entanglement,poboiko2023theory,turkeshi2022enhanced,yokomizo2024measurementinduced}, long-range entangled state preparation~\cite{lu2022measurement,tantivasadakarn2022longrange,tantivasadakarn2023hierarchy,verresen2022efficiently,zhu2022nishimori,lee2022decoding,smith2024constantdepth}, measurement-enhanced entanglement~\cite{cheng2024universal,lin2023probing,zhang2024nonlocal,negari2024measurementinduced}, and critical states under measurements or decoherence~\cite{ashida16critical,ashida17pt,dora20,moca21,buchhold2021effective,minoguchi2022continuous,yamamoto22,ma2023exploring,paviglianiti2023enhanced,garratt2023measurements,lee2023quantum,murciano2023measurementaltered,myerson-jain2023decoherence,sun2023new,weinstein2023nonlocality,yang2023entanglement,zou2023channeling,ashida2023systemenvironment}.
In particular, it has been demonstrated in Refs.~\cite{garratt2023measurements,lee2023quantum,murciano2023measurementaltered,myerson-jain2023decoherence,sun2023new,weinstein2023nonlocality,yang2023entanglement,zou2023channeling,ashida2023systemenvironment} that the effects of measurements on critical spin chains can be described by using boundary conformal field theory (CFT)~\cite{cardy1989boundary,affleck1994fermi,MO97,MO06,difrancesco1997conformal}.

Despite these remarkable developments, our understanding of how measurements enable quantum information transfer between many-body states is still in its infancy.
Quantum critical states are particularly interesting in this context since they possess large entanglement originating from long-range correlations~\cite{briegel2001persistent}.
Motivated by this, we consider entanglement swapping in critical quantum spin chains.
There, the entanglement between the chains occurs \emph{without} direct interactions but with measurements, and a number of fundamental questions arise.
Does the swapped entanglement exhibit universality, and if yes, in what sense?
How does it depend on the number of measured qubits?
These questions are directly relevant to recent experiments realizing measurement and control of many-body systems at the single-quantum level~\cite{bluvstein2022quantum,bluvstein2024logical,MA22,hoke2023measurementinduced}.
While the related questions have been recently addressed in Refs.~\cite{cheng2024universal,sala2024quantum}, theoretical understandings of quantum information transfer between critical chains induced solely by measurements are still lacking.

\begin{figure}[b]
  \includegraphics*[width=0.8\columnwidth,clip]{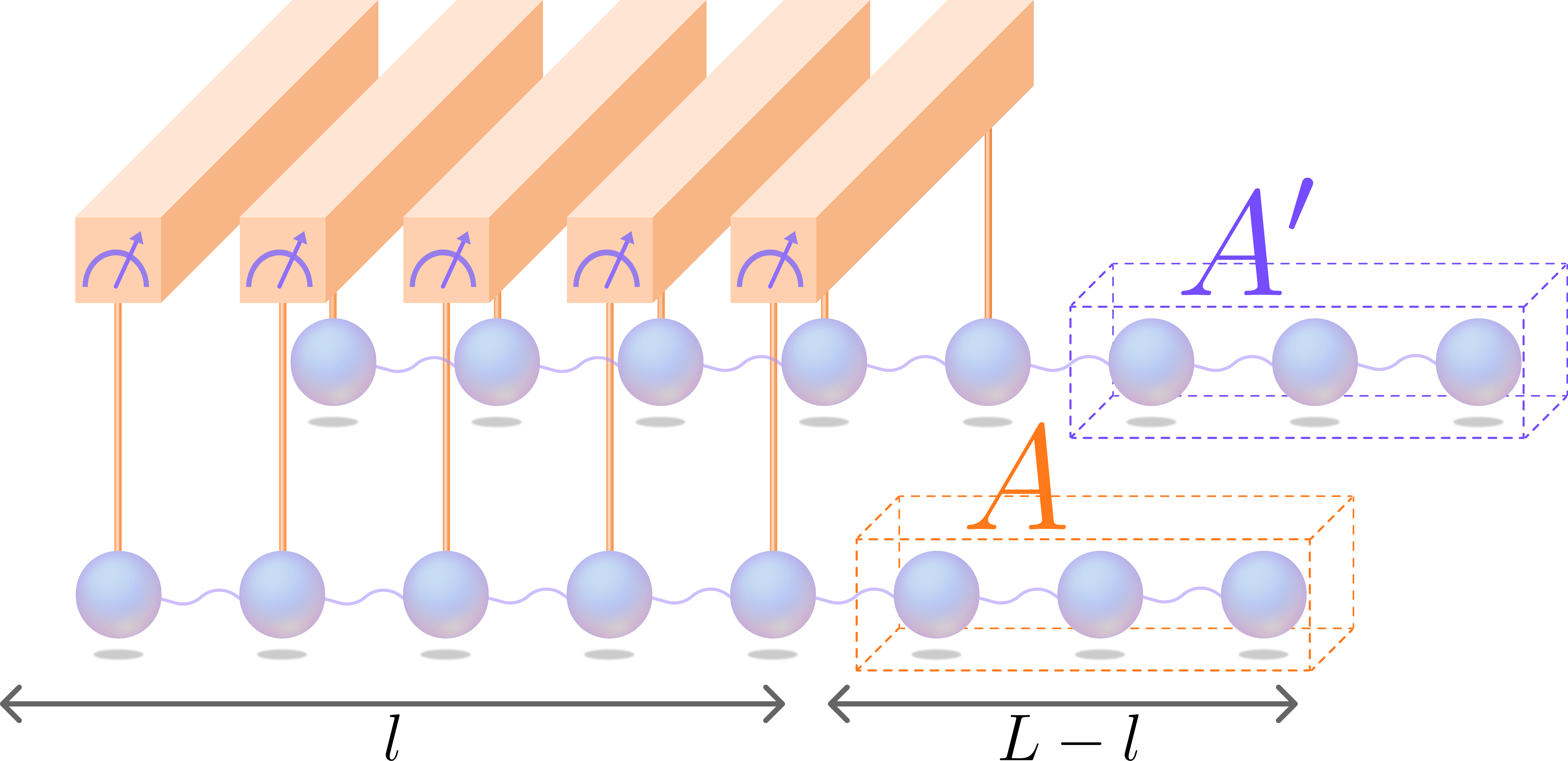}
  \caption{\label{fig:conceptart}Schematic illustration of the setup. The initial state is a pair of independent identical critical ground states. The Bell-state measurements are performed on $l$ qubit pairs, rendering the unmeasured parts $A$ and $A'$ entangled.}
\end{figure}

To address the above questions, we utilize a boundary CFT description and show that the swapped entanglement exhibits the universal logarithmic scaling as a function of the number of measured qubits.
We point out that the Bell-state measurements can impose certain conformal boundary conditions on the time slice of the replicated CFT, leading to the universal coefficient determined by the scaling dimension of the boundary condition changing operator (BCCO).
As a concrete example, we study the critical spin-$\frac{1}{2}$ XXZ chain described by the Tomonaga-Luttinger liquid (TLL)~\cite{tomonaga1950remarks,luttinger1963exactly} and determine the universal coefficient by the boundary CFT analysis.
We numerically verify the field-theoretical results using the tensor-network calculations and discuss possible experimental relevance to Rydberg atom arrays.

\section{Entanglement swapping in critical spin chains}
We consider two initially independent critical spin chains under periodic boundary conditions.
Each of the chains consists of $L$ qubits labeled by $j,j'{\in}\{1,2,{\ldots},L\}$ and is initially in the same critical ground state $\ket{\Psi_0}$.
The entanglement swapping is achieved by performing the interchain Bell-state measurements (Fig.~\ref{fig:conceptart}).
Namely, two qubits across the initially independent chains are measured in the Bell basis, $\ket*{\mathrm{Bell}^{b_1b_2}}{=}(\ket*{0b_1}{+}(-1)^{b_2}\ket*{1\overline{b}_1})/\sqrt{2}$, which is labeled by the four binary numbers $b_1b_2\,{\in}\,\{0,1\}^{2}$.
The projection operator on the Bell basis in qubits $j,j'$ is denoted by $P^{b_1b_2}_{jj'} {=} |\mathrm{Bell}^{b_1b_2}_{jj'}\rangle\langle\mathrm{Bell}^{b_1b_2}_{jj'}|$.
After obtaining measurement outcomes $\vec{m}\,{=}\,(m_1,m_2,\ldots,m_l)\;(m_j\,{\in}\,\qty{0,1}^{2})$, the postmeasurement state becomes
\begin{equation}\label{eq:post-measurement-state}
  \ket*{\Psi^{\vec{m}}}
  = \frac{P^{\vec{m}}\ket*{\Psi_0}\!\otimes\!\ket*{\Psi_0}}{\sqrt{p_{\vec{m}}}}
  = \Biggl[\bigotimes_{j=j'=1}^{l}\!\ket*{\mathrm{Bell}^{m_j}_{jj'}}\Biggr]\otimes\!\ket*{\Psi^{\vec{m}}_{AA'}},
\end{equation}
where $P^{\vec{m}} = \prod_{j=j'=1}^{l}P_{jj'}^{m_j}$ is the projection operator corresponding to the outcome $\vec{m}$, $p_{\vec{m}}\,{=}\,\|P^{\vec{m}}\ket*{\Psi_0}\!\otimes\!\ket*{\Psi_0}\|^2$ is the Born probability, and the unmeasured region in each chain is denoted by $A,A'$, respectively (cf.\ Fig.~\ref{fig:conceptart}).
The second expression in Eq.~\eqref{eq:post-measurement-state} follows from the fact that the projective measurements separate the measured qubits from the rest, rendering the unmeasured part a pure state $\ket*{\Psi^{\vec{m}}_{AA'}}$.

The key point is that the postmeasurement state $\ket*{\Psi^{\vec{m}}_{AA'}}$ now possesses the interchain entanglement between $A$ and $A'$, which is generated solely by the measurement.
This measurement-induced entanglement, which we refer to as the swapped entanglement, can be quantified by the following entanglement entropy (EE),
\begin{equation}\label{eq:swapped-entanglement-EE}
  S^{\vec{m}}_{A} = -\Tr[\rho^{\vec{m}}_{A} \ln \rho^{\vec{m}}_{A}],
\end{equation}
where $\rho_A^{\vec{m}}\,{=}\,\Tr_{A'}[\ketbra*{\Psi^{\vec{m}}_{AA'}}{\Psi^{\vec{m}}_{AA'}}]$ is the reduced density matrix on $A$.
For the sake of later convenience, we write $S^{\vec{m}}_A\,{=}\,S^{\vec{\mu}}_A$ in the case of uniform measurement outcomes where the outcomes for all the pairs coincide, i.e., $\vec{m}\,{=}\,\vec{\mu}\,{=}\,(\mu,\ldots,\mu)\;(\mu\,{\in}\,\{0,1\}^{2})$.
While $S^{\vec{m}}_A$ describes the entanglement properties specific to a measurement outcome $\vec{m}$, we will also analyze the following averaged EE as a quantity representing the typical behavior of the swapped entanglement:
\begin{equation}\label{eq:swapped-entanglement-average}
  \overline{S}_A = \sum_{\vec{m}} p_{\vec{m}} S^{\vec{m}}_{A}.
\end{equation}

\section{General field-theoretical approach}
The universality in the swapped entanglement can be understood by developing the CFT approach.
To this end, we employ the Euclidean path-integral representation and express the two spin chains by a two-component ($1{+}1$)-dimensional field $\bm{\phi}=(\phi,\phi')$, whose action is defined on a two-dimensional sheet.
The effect of the Bell-state measurements can be expressed as an interlayer line defect in this action.
We then use the replica trick to calculate the swapped entanglement from the partition functions of the multicomponent field theory with the line defect.
As a result, we obtain a general form of the swapped entanglement, which shows the universal logarithmic scaling with the number of measurements.

\subsection{Boundary condition imposed by measurements}

\begin{figure}[tb]
  \centering
  \includegraphics[width=0.85\linewidth, clip]{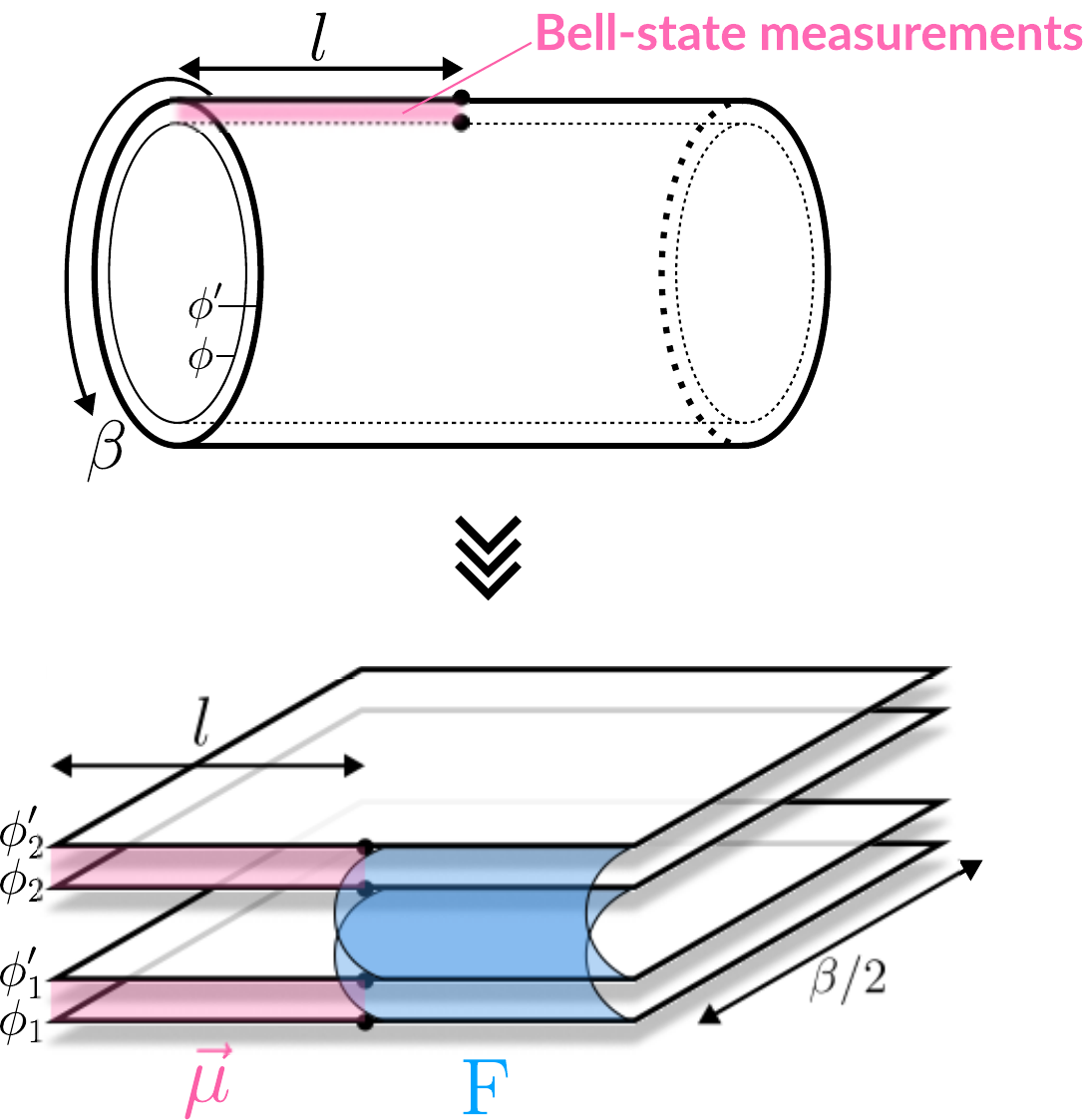}
  \caption{Schematic representation of the boundary CFT approach to calculate the Born probability $p_{\vec{\mu}}$.
    The upper figure shows the Bell-state measurements imposing a interlayer line defect on the torus of size $L\times\beta$ with fields $\phi$ and $\phi'$.
    The figure only illustrates periodicity in the imaginary-time direction.
    The lower figure illustrates the folding into a cylinder of circumference $L$, where the line defect becomes a boundary condition $\vec{\mu}$ at $\tau=0$ in the region $0\leq x\leq l$, while the artificial boundary condition $\mathrm{F}$ is imposed on the rest of the boundary.
  }
  \label{fig:born_probability}
\end{figure}

The initial state $\ket*{\Psi_0}$ is written as the Gibbs state of a critical Hamiltonian $H$:
\begin{equation}
  \rho_0 = \ketbra{\Psi_0}{\Psi_0} = \lim_{\beta\to\infty}\frac{e^{-\beta H}}{\Tr e^{-\beta H}}.
\end{equation}
Hereafter, we consider the ground state and the inverse temprature $\beta$ is understood to be taken infinitely large.
In the path-integral formalism, the expectation value of an observable $A$ in the two spin chains is given as
\begin{equation}
  \langle A \rangle = \Tr[A \;\rho_0\otimes \rho_0] = \frac{1}{Z}\int\!\mathcal{D}\bm{\phi}\; A[\bm{\phi}]\, e^{-\mathcal{S}[\bm{\phi}]}.
\end{equation}
Here, $Z$ is the partition function,
\begin{equation}
  Z = \Tr e^{-\beta H}\!\otimes\! e^{-\beta H} = \int\!\mathcal{D}\bm{\phi}\; e^{-\mathcal{S}[\bm{\phi}]},
\end{equation}
and $\mathcal{S}[\bm{\phi}]$ is the Euclidean action defined on a two-dimensional sheet, which is simply the sum of CFT for each field $\phi$ and $\phi'$.
When the Hamiltonian $H$ is defined on a periodic chain of length $L$, the theory is defined on a torus of size $L\times \beta$ with spatial coordinate $x$ and imaginary time $\tau$.

Now, let us consider the Bell-state measurements on the two spin chains.
The Born probability $p_{\vec{m}}$ is given by
\begin{align}
  p_{\vec{m}} = \Tr[P^{\vec{m}}\;\rho_0\otimes\rho_0]
   & = \frac{1}{Z} \int\!\mathcal{D}\bm{\phi}\; P^{\vec{m}}\,e^{-\mathcal{S}[\bm{\phi}]}\notag            \\
   & = \frac{1}{Z} \int\!\mathcal{D}\bm{\phi}\; e^{-\mathcal{S}[\bm{\phi}]-\delta\mathcal{S}[\bm{\phi}]}.
\end{align}
Here, we assume that the operator $P^{\vec{m}}$ is independent of imaginary time $\tau$, i.e., diagonal in $\bm{\phi}$, and rewrite the projection operator as a boundary perturbation $P^{\vec{m}}=e^{-\delta\mathcal{S}[\bm{\phi}]}$ at a time slice $\tau=0$.
Since the eigenvalue of the projection operator takes either one or zero, the coupling constant in the perturbation becomes infinitely large, leading to a line defect at a time slice on the torus.

For uniform measurement outcomes $\vec{m}=\vec{\mu}$, the Born probability $p_{\vec{\mu}}$ is obtained as a ratio between the partition functions with and without the measurement-induced line defect in the spatial region $0\leq x\leq l$:
\begin{equation}
  p_{\vec{\mu}} = \frac{Z^{\vec{\mu}}}{Z} = \frac{\int\!\mathcal{D}\bm{\phi}\; e^{-\mathcal{S}[\bm{\phi}]-\delta\mathcal{S}[\bm{\phi}]}}{\int\!\mathcal{D}\bm{\phi}\;e^{-\mathcal{S}[\bm{\phi}]}}.
\end{equation}
In this case, the projection operator can be calculated as follows:
\begin{align}
  P_{jj'}^{b_1b_2}
   & = \frac{I+(-1)^{b_1}\sigma_j^z\sigma_{j'}^z}{2}\frac{I+(-1)^{b_2}\sigma_j^x\sigma_{j'}^x}{2}\notag                                                \\
   & = \lim_{g\to\infty}\frac{1}{4\cosh^2g}e^{g[(-1)^{b_1}\sigma_j^z\sigma_{j'}^z +(-1)^{b_2}\sigma_j^x\sigma_{j'}^x]},                                \\
  P^{\vec{\mu}}
   & = \prod_{j=j'=1}^{l} P^{b_1b_2}_{jj'}\notag                                                                                                       \\
   & \propto \lim_{g\to\infty}\exp g\int_0^{l}dx\qty[(-1)^{b_1}\sigma^z\sigma^z + (-1)^{b_2}\sigma^x\sigma^x].\label{eq:boundary-perturbation-general}
\end{align}
Here, $\mu=b_1b_2$ and $\sigma_j^z,\sigma_j^x$ are the Pauli operators at site $j$.
The Pauli operators in the last line is understood to be expressed by the field operator $\bm{\phi}$ describing the long-wavelength region of the theory.
The effect of the Bell-state measurements at the IR limit is determined by the renormalization group (RG) flow of the following boundary perturbation~\footnote{Since the coupling constant $g$ tends to infinity in the present problem, we must be careful using perturbative approaches, e.g., irrelevant terms may also play a role.}:
\begin{equation}
  \delta\mathcal{S}[\bm{\phi}] = -g\!\int_0^\beta \!\!d\tau\!\int_0^l \!dx\,\delta(\tau)\qty[(-1)^{b_1}\sigma^z\sigma^z + (-1)^{b_2}\sigma^x\sigma^x].
\end{equation}
Let us symbolically write the resulting field configuration realized at the line defect as $\bm{\phi}=\bm{\phi}^{\vec{\mu}}$.

It is often convenient to employ the folding trick and rewrite the partition function $Z^{\vec{\mu}}$ as a path integral of a doubled number of fields (denoted $\bm{\phi}_1,\bm{\phi}_2$), defined on a cylinder of length $\beta/2$ and circumference $L$.
As a consequence, the line defect becomes a \textit{boundary condition} in the region $0\leq x\leq l$ at $\tau=0$, and an artificial boundary condition is imposed at the rest of the boundary (Fig.~\ref{fig:born_probability}).
Since we consider the ground state, the limit $\beta\to\infty$, the effect of the boundary condition at $\tau=\beta/2$ can be ignored.
Therefore, the partition function $Z^{\vec{\mu}}$ can be calculated as a path-integral on a cylinder with the following boundary condition at $\tau=0$:
\begin{align}\label{eq:born-boundary-condition}
  \vec{\mu}  & : \; \bm{\phi}_1(x,0) = \bm{\phi}_2(x,0) = \bm{\phi}^{\vec{\mu}}(x) \quad (0\leq x\leq l), \\
  \mathrm{F} & : \;\bm{\phi}_1(x,0) - \bm{\phi}_2(x,0)=0 \quad (l< x\leq L).
\end{align}

\subsection{Replica trick}

\begin{figure}[tb]
  \centering
  \includegraphics[width=0.8\linewidth, clip]{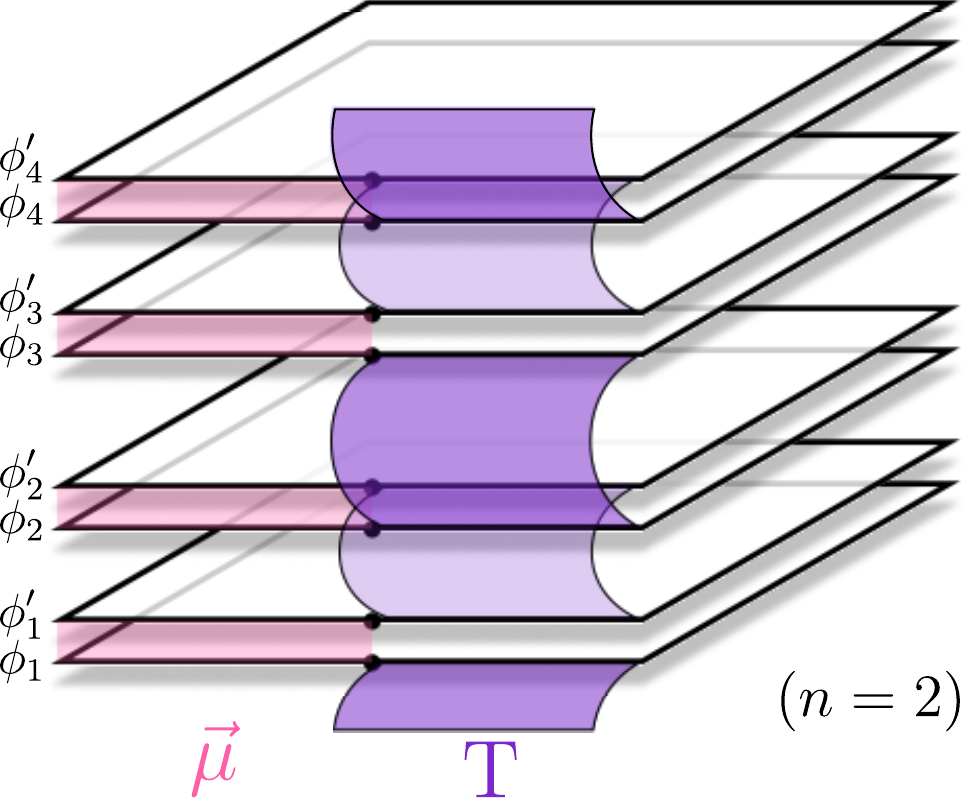}
  \caption{Illustration of the partition function $Z^{\vec{\mu}}_n(A)$ for $n=2$.
    The boundary condition $\vec{\mu}$ is imposed on the measured region $0\leq x\leq l$, while boundary condition $\mathrm{T}$ is imposed on the unmeasured region $l< x\leq L$.
    The boundary condition $\mathrm{T}$ connects the replicas to the next in subsystem $A$, but take the trace in subsystem $A'$ by connecting with itself.}
  \label{fig:replica_partition_function}
\end{figure}

To analyze the swapped EE, we use the replica trick~\cite{HOLZHEY1994443,calabrese2004entanglement,calabrese2009entanglement,calabrese2012entanglement,calabrese2013entanglement} and evaluate the EE by taking $n\to 1$ in $-\partial_n\Tr[(\rho^{\vec{\mu}}_A)^n]$.
To calculate the latter quantity, we need to replicate the field $n$-times and sew them in region $A$, while the boundary interaction is induced by the measurement on the spatial region in $0\leq x\leq l$.
This can be expressed by the partition functions with boundaries through the following:
\begin{align}
  \Tr_A[(\rho_A^{\vec{\mu}})^n]
   & = \frac{1}{(p_{\vec{\mu}})^n}\Tr[(P^{\vec{\mu}}\rho_0\!\otimes\!\rho_0P^{\vec{\mu}})^{\otimes n}\tau_{n,A}]\notag \\
   & = \qty(\frac{Z}{Z^{\vec{\mu}}})^n \frac{Z_n^{\vec{\mu}}(A)}{Z^n} = \frac{Z_n^{\vec{\mu}}(A)}{(Z^{\vec{\mu}})^n}.
\end{align}
Here, $\tau_{n,A}$ applies a forward permutation of the replicas in subsystem $A$ and acts as an identity in the rest of the system.
After the folding trick, the partition function $Z_n^{\vec{\mu}}(A)$ can be calculated as a path-integral on a cylinder with the following boundary condition at $\tau=0$:
\begin{align}\label{eq:replica-boundary-condition}
  \vec{\mu}  & : \; \bm{\phi}_{2i-1}(x,0) = \bm{\phi}_{2i}(x,0) =\bm{\phi}^{\vec{\mu}}(x)\quad (0\leq x\leq l), \\
  \mathrm{T} & : \left\{
  \begin{aligned}
     & \phi'_{2i-1}(x,0)  = \phi'_{2i}(x,0)  &  & (l< x\leq L) \\
     & \phi_{2i}(x,0)     = \phi_{2i+1}(x,0) &  & (l< x\leq L)
  \end{aligned}
  \right..
\end{align}
Here, $i=1,2,\ldots,n$ represents the label of the replicas, and $\phi_{2n+1}=\phi_1$.
The number of the fields is $4n$ due to the folding trick.
The boundary condition $\mathrm{T}$ contains the sewing condition, imposed by the action of $\tau_{n,A}$ in the trace (Fig.~\ref{fig:replica_partition_function}).
In summary, the swapped EE is obtained from the partition functions $Z^{\vec{\mu}}$ and $Z^{\vec{\mu}}_n(A)$, where both of them have two different boundary conditions between the regions $0\leq x\leq l$ and $l< x\leq L$.

\subsection{Boundary condition changing operator}
In general, a boundary of a CFT in the IR limit is described by a conformally invariant boundary condition which is determined by the boundary condition imposed at the microscopic scale, e.g., the Bell-state measurements in the present problem.
When two neighboring segments of the boundary are subject to conformally invariant boundary conditions, we can interpret it as an insertion of BCCO at the point separating the two segments.
In the present case, $Z^{\vec{\mu}}_n(A)$ and $Z^{\vec{\mu}}$ have the BCCOs from $\vec{\mu}$ to $\mathrm{T}$ and $\vec{\mu}$ to $\mathrm{F}$ at $x=0,l$, respectively.
Thus, if the boundary condition $\vec{\mu}$ is conformally invariant, then the partition functions can be expressed as two-point correlation functions of the BCCOs,
\begin{align}
  Z^{\vec{\mu}}_n(A)
   & =  \ev{\mathcal{B}_{\mathrm{T}\vert\vec{\mu}}(l)\mathcal{B}_{\vec{\mu}\vert\mathrm{T}}(0)}, \\
  Z^{\vec{\mu}}
   & =  \ev{\mathcal{B}_{\mathrm{F}\vert\vec{\mu}}(l)\mathcal{B}_{\vec{\mu}\vert\mathrm{F}}(0)},
\end{align}
where $\mathcal{B}_{b\vert a}(x)$ is a BCCO that changes the boundary condition from $a$ to $b$ at point $x$.

In CFT, two-point correlation functions are determined only by the scaling dimensions of the operators.
If we write the scaling dimension of the BCCO $\mathcal{B}_{b\vert a}(x)$ as $\Delta_{b\vert a}$ and evaluate the correlation function on a cylinder of circumference $L$, then the partition functions behave as
\begin{align}
  Z^{\vec{\mu}}_n(A)/Z^n & = e^{-n\epsilon l}\, l_c^{-2\Delta_{\mathrm{T}\vert\vec{\mu}}},\label{eq:replica-behavior} \\
  Z^{\vec{\mu}}/Z        & = e^{-\epsilon l}\, l_c^{-2\Delta_{\mathrm{F}\vert\vec{\mu}}}\label{eq:born-behavior},
\end{align}
where $\epsilon$ is a nonuniversal line-energy density and $l_c=(L/\pi)\sin(\pi l/L)$ is the chord length.
We note that the nonuniversal \textit{bulk} energy cancels due to the normalization by the original partition function $Z$.

\subsection{Result: general form of the swapped EE}
From the expressions in Eqs.~\eqref{eq:replica-behavior} and~\eqref{eq:born-behavior}, we obtain the general form of the swapped EE as follows:
\begin{equation}\label{eq:uniform-EE}
  S^{\vec{\mu}}_{A}
  = -\lim_{n\to1}\frac{\partial}{\partial n}\frac{Z^{\vec{\mu}}_n(A)}{(Z^{\vec{\mu}})^n}
  \sim c_{\vec{\mu}} \ln\qty[\frac{L}{\pi}\sin\frac{\pi l}{L}].
\end{equation}
Here, we introduce the universal coefficient,
\begin{equation}\label{eq:coefficient}
  c_{\vec{\mu}}=\lim_{n\to1}\frac{2(\Delta_{\mathrm{T}\vert\vec{\mu}}-n\Delta_{\mathrm{F}\vert\vec{\mu}})}{n-1},
\end{equation}
which is determined by the scaling dimensions of the BCCOs.
The result is universal in the sense that the scaling dimensions of the BCCOs are characterized by the IR limit of the theory, independent of microscopic details of the models.
Under a certain assumption for the boundary condition $\vec{\mu}$, this coefficient can be shown to be $c_{\vec{\mu}}=c/6$ for $\mu=00$, which only depends on the central charge $c$ on the underlying CFT (see Appendix~\ref{appendix:general} for derivation).

The above argument can be extended to the case of a general measurement outcome $\vec{m}$.
Let $x_i$ denote the boundary condition changing points for $\vec{m}$.
Suppose that the domains between those points, where the outcomes are the same, extend over the regions longer than the lattice constant so that the continuum description remains valid.
The path-integral representation of $S^{\vec{m}}_A$ can then be obtained as a multipoint correlation function of the BCCOs as follows:
\begin{equation}\label{eq:general-EE}
  \Tr[(\rho^{\vec{m}}_A)^n]
  = \frac{\ev{\prod_{i}\mathcal{B}_{b_{i+1}\vert b_{i}}(x_i)}}{\ev{\prod_{i}\mathcal{B}_{a_{i+1}\vert a_i}(x_i)}^n}.
\end{equation}
Here, $b_i$ is either the sewing condition ($\mathrm{T}$) or a boundary condition induced by any of the Bell bases ($\vec{\mu}$), and $a_i$ is the same as in $b_i$ except that $\mathrm{T}$ should be replaced by $\mathrm{F}$.

We may also express the averaged swapped EE $\overline{S}_A$ by employing the replica trick.
To see this, we introduce the unnormalized density matrices $\tilde{\rho}^{\vec{m}} {=}\, p_{\vec{m}}\ketbra*{\Psi^{\vec{m}}}{\Psi^{\vec{m}}}$ and $\tilde{\rho}^{\vec{m}}_A{=}\,p_{\vec{m}}\rho^{\vec{m}}_A$.
The averaged one then reads
\begin{equation}\label{eq:ensemble-EE}
  \overline{S}_A
  = \lim_{\substack{n\to 1\\ s\to 0}} \sum_{\vec{m}} \frac{\Tr[\tilde{\rho}^{\vec{m}}]\Tr[(\tilde{\rho}^{\vec{m}}_A)^n]^s - \Tr[\tilde{\rho}^{\vec{m}}]^{ns+1}}{s(1-n)}.
\end{equation}
This expression corresponds to a path integral of $ns{+}1$ replicas with certain boundary interactions.
When the boundary conditions are conformally invariant, the averaged swapped EE $\overline{S}_A$ also obeys the universal logarithmic scaling as demonstrated below.

\section{Entanglement swapping in critical XXZ chains}
As a concrete example, we consider the entanglement swapping in the two copies of the spin-$\frac{1}{2}$ critical XXZ chain,
\begin{equation}\label{eq:xxz-hamiltonian}
  H = \sum_{j=1}^{L} (\sigma_j^x\sigma_{j+1}^x + \sigma_j^y\sigma_{j+1}^y + \Delta \sigma_j^z\sigma_{j+1}^z),\;\; \Delta\in(-1,1],
\end{equation}
whose effective field theory is given by a $c\,{=}\,1$ CFT with the TLL parameter  $K\,{=}\,\pi/(2\pi \,{-}\, 2\cos^{-1}\Delta)$~\cite{giamarchi2003quantum}.
We recall that the TLL is described by a boson field $\phi$ and its dual $\theta$, which are compactified as $\phi\,\sim\,\phi\,{+}\,\pi n,\theta\,\sim\,\theta\,{+}\,2\pi m\;(n,m\,{\in}\,\mathbb{Z})$.
These fields can be related to the $j$th qubit $\sigma_j^\alpha\;(\alpha\,{=}\,x,y,z)$ by
\begin{align}
  \sigma_j^z
   & \simeq \frac{2}{\pi}\partial_x\phi + \frac{(-1)^j}{\pi\gamma}\cos(2\phi),\label{eq:bosonization-relation_z}   \\
  \sigma_j^x \pm i\sigma_j^y
   & \simeq \frac{e^{\pm i\theta}}{\sqrt{\pi\gamma}}\qty[(-1)^j + \cos(2\phi)],\label{eq:bosonization-relation_xy}
\end{align}
where $\gamma$ is a nonuniversal constant of order unity.
The bosonic fields corresponding to each of the two chains are denoted by $\phi,\phi'$, respectively, and we write their duals as $\theta,\theta'$.

\subsection{Identifying the boundary conditions}
To determine the universal coefficient $c_{\vec{\mu}}$ in Eq.~\eqref{eq:uniform-EE}, we first need to identify the conformal boundary condition $\vec{\mu}$ induced by the measurement operator $P^{\vec{\mu}}$.
Using the bosonization relation [Eqs.~\eqref{eq:bosonization-relation_z} and~\eqref{eq:bosonization-relation_xy}], we can represent the Bell-state measurement as a boundary perturbation $P^{\vec{\mu}}\,{=}\,e^{-\delta\mathcal{S}}$ with the boundary action $\delta\mathcal{S}$ being given by
\begin{multline}\label{eq:measurementOp-bosonized}
  \delta\mathcal{S} =-\lim_{g\to\infty} g\int_0^\beta \!\!d\tau\!\int_0^l \!dx\,
  \delta(\tau)\Bigl[\; (-1)^{b_2}\cos\theta_{b_1}\\
    + (-1)^{b_1}\frac{\cos(2\phi_+) + \cos(2\phi_-)}{2\pi\gamma}\Bigr],
\end{multline}
where $\phi_{\pm}\,{=}\,\phi\pm\phi'$, $\theta_{\pm}\,{=}\,\theta\pm\theta'$, and $\theta_{b_1=0/1}$ represents $\theta_{+/-}$, respectively.
The resulting boundary condition is the configuration that minimizes $\delta\mathcal{S}$.
Consequently, there are four possible boundary conditions depending on the measurement outcomes, where the boson fields are locked at the boundary such that $(\phi_-,\theta_+){=}(0,0)$, $(\phi_-,\theta_+){=}(0,\pi)$, $(\phi_+,\theta_-){=}(\pi/2,0)$, and $(\phi_+,\theta_-){=}(\pi/2,\pi)$ for $\mu\,{=}\,00,01,10,11$, respectively.
These mixed Dirichlet-Neumann conditions are known to be conformally invariant~\cite{MO06}.
We note that the configuration corresponding to $\mu{=}10$ ($\mu{=}11$) has been also known to represent the rung singlet$^\ast$ (rung singlet) phase in the spin-$\frac{1}{2}$ XXZ ladder~\cite{ogino2021symmetryprotected,ogino2022groundstate,mondal2023symmetryenriched,fontaine2024symmetry}.

In the present case, the swapped EE $S_A^{\vec{\mu}}$ takes the same value for all four Bell bases $\vec{\mu}$, because the Bell bases relate to each other by a qubitwise unitary operator, and the initial state $\ket*{\Psi_0}\otimes\!\ket*{\Psi_0}$ is invariant under this unitary operation.
Thus, it suffices to focus on a particular outcome which we choose to be $\mu=00$, i.e., the boundary condition $\vec{\mu}:\;\phi = \phi'$.

\subsection{Calculating the scaling dimensions of the BCCOs}

\begin{figure}[tb]
  \centering
  \includegraphics[width=0.7\linewidth, clip]{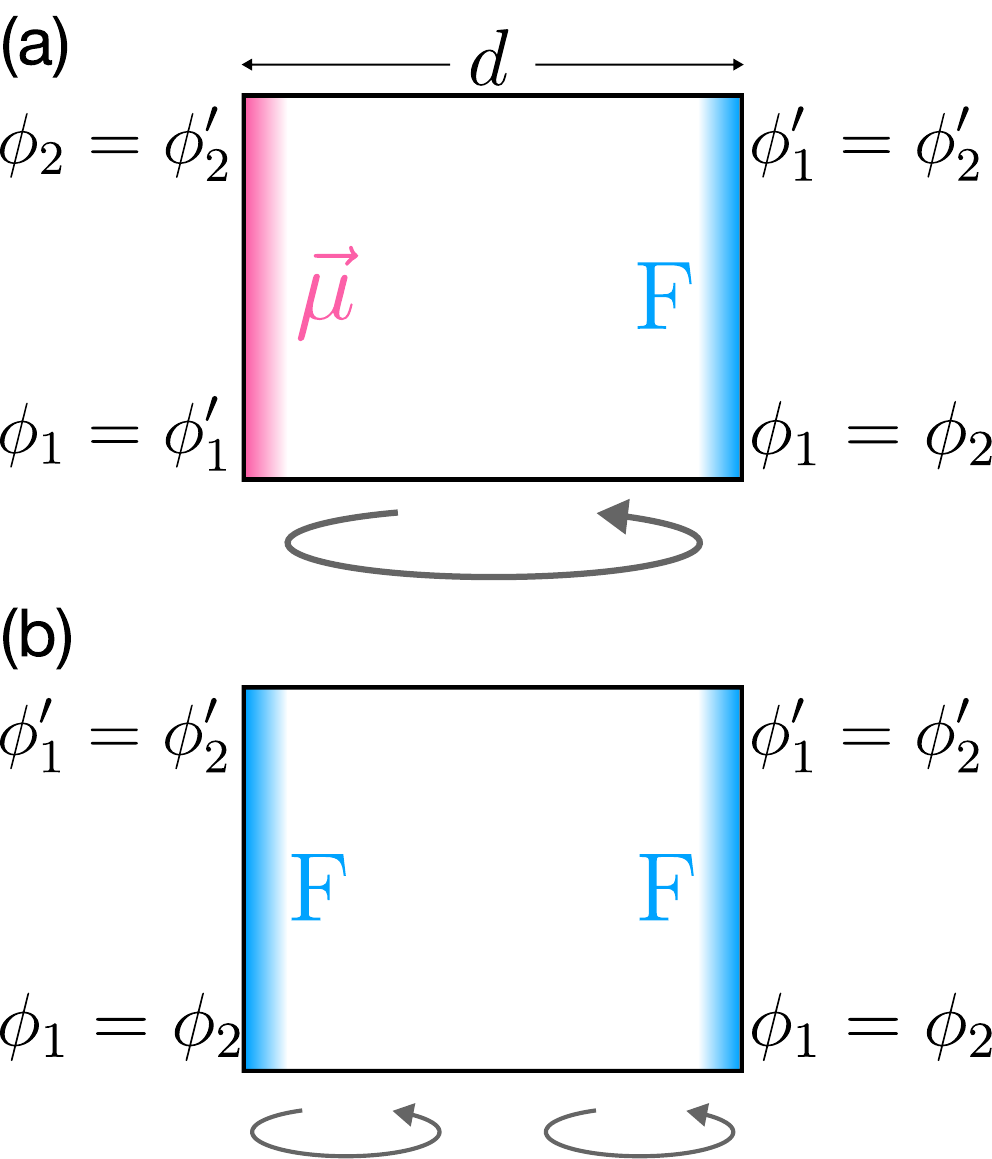}
  \caption{(a)~The strip geometry with boundary condition $\vec{\mu}$ on the left, and $\mathrm{F}$ on the right.
    Time flows in the vertical direction, and the ground-state energy of the corresponding Hamiltonian is $E_{\mathrm{F}\vec{\mu}}^0(d)$.
    We can see that the four replicas form a single loop, and the strip geometry effectively becomes a cylinder.
    (b)~The strip geometry with boundary condition $\mathrm{F}$ on both sides.
    The four replicas form a pair of loops.}
  \label{fig:gs_energy_f}
\end{figure}

We next derive the scaling dimensions of the BCCOs $\mathcal{B}_{\mathrm{T}\vert\vec{\mu}}$ and $\mathcal{B}_{\mathrm{F}\vert\vec{\mu}}$.
For this purpose, we note that the scaling dimension $\Delta_{b\vert a}$ of a BCCO $\mathcal{B}_{b\vert a}$ can be related to the ground-state energy of the corresponding CFT on a strip geometry where boundary conditions $a$ and $b$ are imposed on each side~\cite{affleck1997boundary}.
Specifically, we use the following relation:
\begin{equation}
  \frac{\pi\Delta_{b\vert a}}{d}  = E_{ab}^0(d) - E_{aa}^0(d),
\end{equation}
where $d$ is the width of the strip geometry and $E_{ab}^0(d)\;(E_{aa}^0(d))$ is the ground-state energy of the CFT on the strip geometry with boundary conditions $a,b\;(a,a)$ on both ends (see Appendix~\ref{appendix:bcco} for derivation).
Determining the ground-state energy for a general geometry is not an easy task.
However, if the CFT is on a cylinder, then the ground-state energy is known to be the Casimir energy:
\begin{equation}
  E^0_{\mathrm{cyl}} = -\frac{\pi}{6}\frac{c}{L_{\mathrm{cyl}}}.
\end{equation}
Here, $L_{\mathrm{cyl}}$ is the circumference of the cylinder, and $c$ is the central charge of the CFT.

Let us start with $Z^{\vec{\mu}}$, the partition function of the four-component field with boundary conditions $\vec{\mu}$ and $\mathrm{F}$.
The ground-state energy $E_{\mathrm{F}\vec{\mu}}^0(d)$ is equivalent to the Casimir energy of a single-component boson field theory on a cylinder of circumference $4d$:
\begin{equation}
  E_{\mathrm{F}\vec{\mu}}^0(d) = -\frac{\pi}{6}\frac{1}{4d} = -\frac{\pi}{24 d}.
\end{equation}
This can be seen from the illustration in Fig.~\ref{fig:gs_energy_f}, where the fields effectively form a closed loop along the strip geometry, and becomes a cylinder of a single-component field.
The same applies to the ground-state energy $E_{\mathrm{F}\mathrm{F}}^0(d)$, which is equivalent to twice the Casimir energy of a single-component boson field theory on a cylinder of circumference $2d$, namely,
\begin{equation}
  E_{\mathrm{F}\mathrm{F}}^0(d) = -\frac{\pi}{6}\frac{1}{2d}\times 2 = -\frac{\pi}{6d}.
\end{equation}
Thus, the scaling dimension $\Delta_{\mathrm{F}\vert\vec{\mu}}$ reads
\begin{align}\label{eq:scaling_dimension_f}
  \Delta_{\mathrm{F}\vert\vec{\mu}}
   & = \frac{d}{\pi}\qty(E_{\mathrm{F}\vec{\mu}}^0(d) - E_{\mathrm{F}\mathrm{F}}^0(d))\notag \\
   & = \frac{1}{8}.
\end{align}

\begin{figure}[tb]
  \centering
  \includegraphics[width=0.7\linewidth, clip]{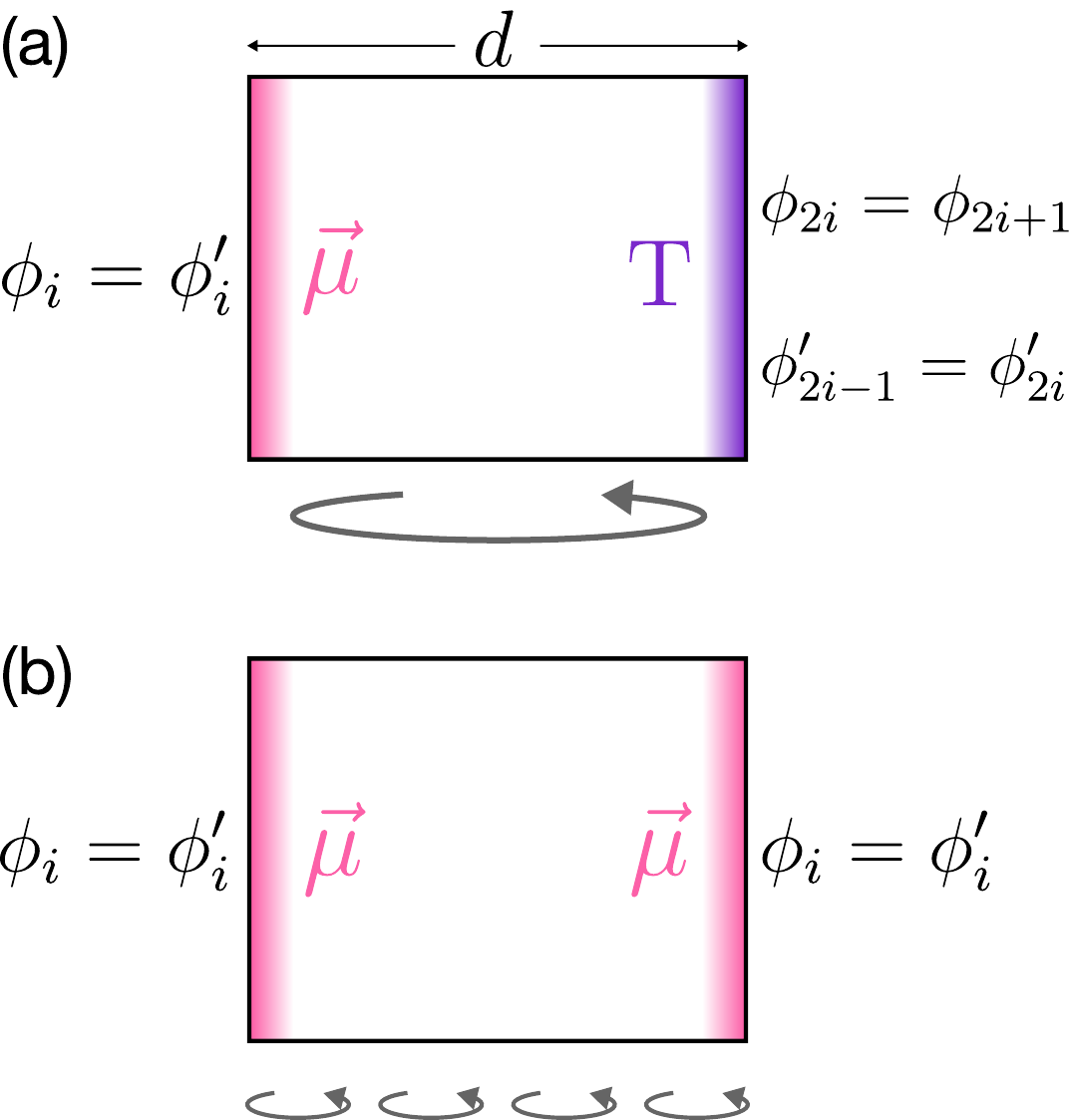}
  \caption{(a)~The strip geometry with boundary condition $\vec{\mu}$ on the left and $\mathrm{T}$ on the right.
    The $4n$ replicas form a single loop.
    (b)~The strip geometry with boundary condition $\vec{\mu}$ on both sides.
    The $4n$ replicas form four loops.}
  \label{fig:gs_energy_t}
\end{figure}

We will now calculate $\Delta_{\mathrm{T}\vert\vec{\mu}}$, the scaling dimension of the BCCO $\mathcal{B}_{\mathrm{T}\vert\vec{\mu}}$ used to express the partition function of the replicas $Z_n^{\vec{\mu}}(A)$.
Again, the strip geometry with the boundary condition $\mathrm{T}$ and $\vec{\mu}$ on each side effectively becomes a cylinder (Fig.~\ref{fig:gs_energy_t}).
Since the $4n$ components form a single loop, the circumference of the cylinder is $4dn$.
Thus,
\begin{equation}
  E_{\mathrm{T}\vec{\mu}}^0(d) = -\frac{\pi}{6}\frac{1}{4dn} = -\frac{\pi}{24 dn}.
\end{equation}
On the other hand, $E_{\vec{\mu}\vec{\mu}}^0(d)$ is given by $2n$ times the ground-state energy of the free-boson CFT on a cylinder of circumference $2d$, since there are $2n$ loops formed by $2n$ pairs of boson fields $\phi_i$ and $\phi'_i$.
\begin{equation}
  E_{\vec{\mu}\vec{\mu}}^0(d) = -\frac{\pi}{6}\frac{1}{2d}\times 2n = -\frac{\pi n}{6d}.
\end{equation}
The scaling dimension is then
\begin{align}\label{eq:scaling_dimension_t}
  \Delta_{\mathrm{T}\vert\vec{\mu}}
   & = \frac{d}{\pi}\qty(E_{\mathrm{T}\vec{\mu}}^0(d) - E_{\vec{\mu}\vec{\mu}}^0(d))\notag \\
   & = \frac{n}{6} - \frac{1}{24n}.
\end{align}

From the results in Eqs.~\eqref{eq:coefficient},~\eqref{eq:scaling_dimension_f}, and~\eqref{eq:scaling_dimension_t}, we obtain the following value for the universal coefficient:
\begin{equation}
  c_{\vec{\mu}} = \frac{1}{6}.
\end{equation}
Recall that the initial entanglement scales logarithmically with the coefficient $1/3$, and this result implies that for the Bell-state measurements with uniform outcomes, half of the initial entanglement is swapped, i.e., the efficiency is $1/2$.

We note that this calculation for the scaling dimensions of BCCOs can be applied to derive the universal logarithmic scaling of the EE in the initial state.
There, the strip geometry also becomes a cylinder, and the scaling dimension can be related to the Casimir energy, which is determined only by the central charge of the CFT.
See Appendix~\ref{appendix:ee} for details.

\begin{figure}[tb]
  \includegraphics*[width=\linewidth,clip]{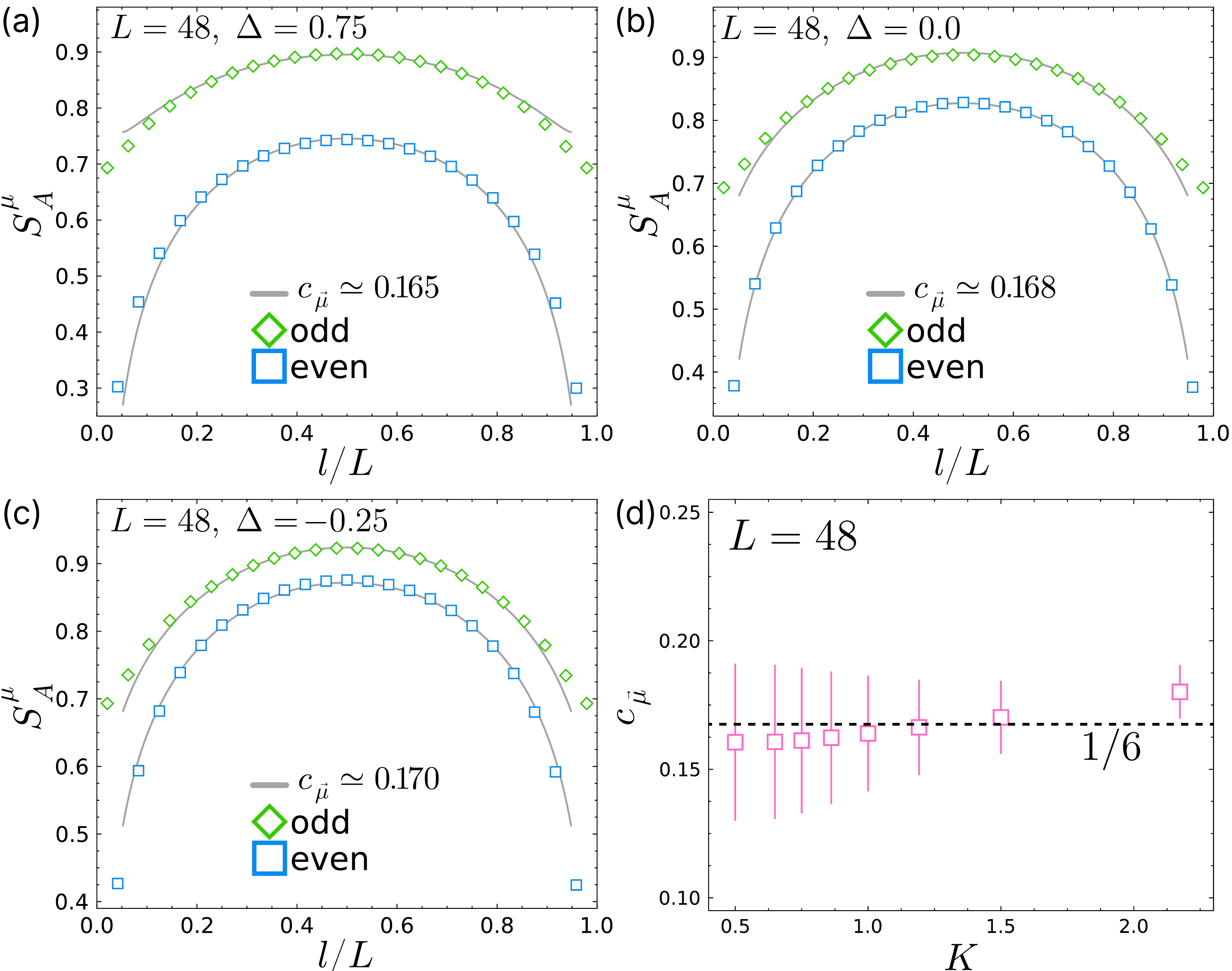}
  \caption{\label{fig:xxz_uniform}[(a)-(c)]~Numerical results of $S^{\vec{\mu}}_A$ at different $\Delta$ plotted as a function of $l/L$.
    Markers are changed for $l$ even and odd.
    The black curves are obtained by fitting the results to Eq.~\eqref{eq:fitting}.
    The fitting parameters at $\Delta\,{=}\,0.75$, $0.0$, and $-0.25$ are $(c_{\vec{\mu}},c_1,c_2)=(0.165,-0.44,0.05)$, $(0.168,-0.24,0.12)$, and $(0.170,-0.15,0.13)$, respectively.
    (d)~The universal coefficient $c_{\vec{\mu}}$ plotted against $K$.
    Error bars indicate variations when the region used for the fitting is varied.
  }
\end{figure}

\subsection{Numerical calculations}
To numerically test the above results, we perform matrix product states (MPS) calculations of the entanglement swapping in the XXZ chains.
Figure~\ref{fig:xxz_uniform}(a)-\ref{fig:xxz_uniform}(c) verifies the predicted logarithmic scaling of $S^{\vec{\mu}}_A$ as a function of the number of measured qubits $l$.
Additionally, we observe an even-odd oscillation depending on $l$.
This parity effect is analogous to what has been found in the entanglement entropy of an XXZ chain with open boundary conditions~\cite{laflorencie2006boundary,sorensen2007quantum,affleck2009entanglement}, indicating that the projective measurements effectively create open ends in the chains.
Following Refs.~\cite{laflorencie2006boundary,sorensen2007quantum,affleck2009entanglement}, we consider the fitting function that includes this oscillation term,
\begin{equation}\label{eq:fitting}
  S^{\vec{\mu}}_A\!=\! c_{\vec{\mu}} \ln\qty[\frac{L}{\pi}\sin(\frac{\pi l}{L})] \!+\! c_1(c_2 + (-1)^l)\qty[\frac{L}{\pi}\sin(\frac{\pi l}{L})]^{-K},
\end{equation}
where $c_1,c_2$ are nonuniversal constants whose values are provided in the caption of Fig.~\ref{fig:xxz_uniform}.
Figure~\ref{fig:xxz_uniform}(d) shows the estimated value of the coefficient $c_{\vec{\mu}}$ at different $K$, which confirms the predicted universal value  $c_{\vec{\mu}}\,{=}\,1/6$ within the error bars.
We note that the data points close to the edges ($l\,{\sim}\, 0,L$) are not included in the fitting since they substantially deviate from the logarithmic behavior due to the short-distance effect.
The error bars in Fig.~\ref{fig:xxz_uniform}~(d) are mainly due to the fact that the estimated values of $c_{\vec{\mu}}$ exhibit relatively large fluctuations depending on whether the number of those excluded data points were even or odd.

\begin{figure}[tb]
  \includegraphics*[width=\columnwidth,clip]{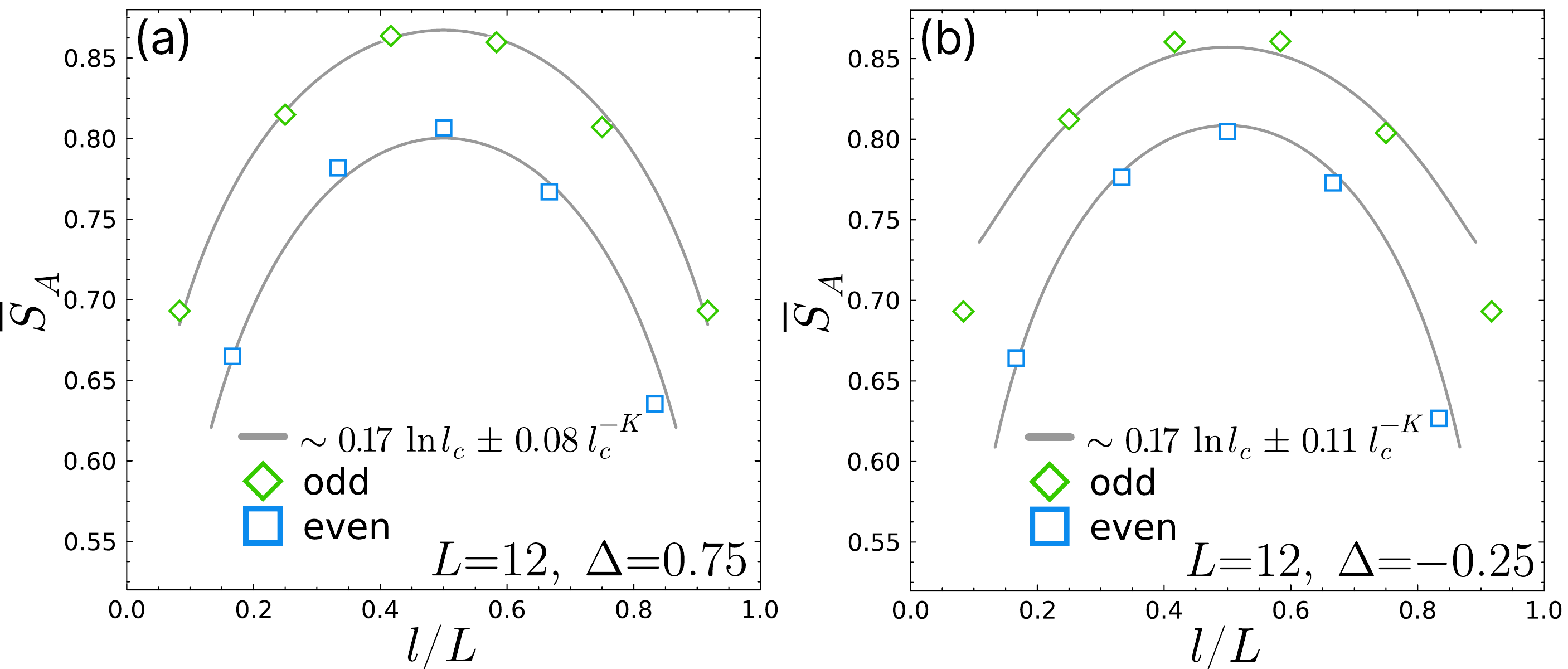}
  \caption{\label{fig:xxz_ensemble}[(a) and (b)]~Numerical results of the averaged swapped entanglement $\overline{S}_A$ at (a)~$\Delta{=}0.75$ and (b)~$\Delta{=}{-}0.25$.
  Fitting curves are shown in the legend as a function of $l_c{=}(L/\pi)\sin(\pi l/L)$.
  Statistical errors are small enough compared to the size of the markers.
  }
\end{figure}

We next numerically evaluate the averaged swapped entanglement $\overline{S}_A$, where the ensemble average is taken over all the possible measurement outcomes [see Eq.~\eqref{eq:swapped-entanglement-average}].
Interestingly, as shown in Fig.~\ref{fig:xxz_ensemble}, we find that $\overline{S}_A$ again exhibits the logarithmic scaling and the parity effect, which are akin to what we find in the uniform outcome cases above.
In particular, the estimated value of the coefficient in the logarithmic term ($0.17$) is close to that of the uniform case $c_{\vec{\mu}}\,{=}\,1/6$.
These results suggest that the measurement-induced boundary interactions in the CFT of $ns{+}1$ replicas in Eq.~\eqref{eq:ensemble-EE} lead to certain conformal boundary conditions, and the scaling dimension of the corresponding BCCO should be the same as in the uniform case (see Appendix~\ref{appendix:averaged} for further discussions).

\section{Experimental accessibility}
Our consideration should be relevant to current experiments.
The key requirements for the entanglement swapping proposed in this paper are the preparation of the critical states, the Bell-state measurements, and the evaluation of the EE.
All these techniques are within reach of current programmable quantum platforms~\cite{ho2019efficient,kalinowski2023nonabelian}, such as Rydberg atom arrays~\cite{bluvstein2022quantum,bluvstein2024logical}.
We note that our setup does not include mid-circuit measurements as in hybrid quantum circuits~\cite{hoke2023measurementinduced}, and the postmeasurement state is pure.
Also, the signature of the universality in the swapped entanglement is appreciable even in a small system with, e.g., 10 qubits.
Thus, an experiment with brute-force postselections and/or quantum state tomography might be feasible.
To circumvent the exponential overhead for a larger number of qubits, it might be useful to employ recent suggestions of mutually unbiased bases quantum state tomography~\cite{koh2023measurementinduced}, MPS tomography~\cite{cramer2010efficient,lanyon2017efficient}, machine-learning approaches~\cite{torlai2018neuralnetwork,carrasquilla2019reconstructing}, classical shadows technique~\cite{huang2020predicting,vermersch2024enhanced}, and classical simulation-assisted schemes~\cite{garratt2023probing,mcginley2024postselectionfree}.

\section{Summary and Discussion}
We have shown that the swapped entanglement in critical spin chains can exhibit the universal logarithmic behavior.
We have provided a boundary CFT prescription for calculating the universal coefficient when the Bell-state measurement outcomes are uniform (cf.\ Eqs.~\eqref{eq:uniform-EE} and~\eqref{eq:coefficient}).
The field-theoretical results have been confirmed in the case of critical XXZ chains through numerical calculations (cf.\ Fig.~\ref{fig:xxz_uniform}).
A similar universal behavior has been observed numerically when the ensemble average is taken over all the measurement outcomes (cf.\ Fig.~\ref{fig:xxz_ensemble}).

There are several intriguing directions for future studies.
First, it merits further study to understand entanglement swapping of critical states in a different universality class, such as the Ising criticality or even for CFTs realized in qudit systems.
For time-reversal symmetric systems with uniform measurement outcomes of $\ket*{\mathrm{Bell}^{00}}$, we found that the swapped entanglement can be related to the entanglement of the initial state.
Thus, the swapped entanglement $S^{\vec{\mu}}_A \; (\mu=00)$ in critical states with time-reversal symmetry depend only on the central charge $c$ of the underlying CFT.
We can also derive this result from CFT under certain assumptions of the boundary condition imposed by the measurements (see Appendix~\ref*{appendix:general}).

Secondly, it is of practical importance to identify what would be the most efficient entanglement swapping protocol.
One can, for instance, optimize a choice of the type/number of the initial chains and/or the measurement basis to maximize the amount of the swapped entanglement.
Our findings suggest that, among all the possible measurements that lead to the conformal boundary conditions, the one with the highest scaling dimension should be the optimal choice.

Lastly, it would be worthwhile to consider a possible extension of our boundary CFT analysis to the problem of quantum state learning~\cite{dallarno2011informational,barratt2022transitions,elben2023randomized,khemani2024learnability}.
While there have been significant advances in our understanding of the effects of measurement backaction on critical states, their implications from a perspective of the learnability of quantum states is largely unexplored.
Our approach allows for a quantitatively accurate characterization of long-distance entanglement properties in a postmeasurement state, which might be useful also in this context. We hope that our study stimulates further studies in these directions.

%%%=========== main text end ===========%%%

%%% acknowledgments %%%
\begin{acknowledgments}
  We are grateful to Yohei Fuji, Shunsuke Furukawa, Marcin Kalinowski, Ruochen Ma, and Shinsei Ryu for valuable discussions.
  M.H. thanks Yuki Koizumi and Kaito Watanabe for helpful comments.
  We thank Akihiro Hokkyo for pointing out the general result for time-reversal symmetric states described in Appendix~\ref{appendix:general}.
  We used the ITensor package~\cite{10.21468/SciPostPhysCodeb.4,10.21468/SciPostPhysCodeb.4-r0.3} for MPS calculations.
  The authors thank the Yukawa Institute for Theoretical Physics (YITP), Kyoto University; discussions during the YITP workshop were useful to complete this work.
  Y.A. acknowledges support from the Japan Society for the Promotion of Science through Grant No.~JP19K23424 and from JST FOREST Program (Grant No.~JPMJFR222U, Japan).
  The work of M.O. was partially supported by the JSPS KAKENHI Grants No.~JP23K25791 and No.~JP24H00946.
\end{acknowledgments}

%%%=========== Appendix ===========%%%

\appendix

\section{\label{appendix:bcco}Scaling dimension of BCCOs}

\begin{figure}[tb]
  \includegraphics*[width=\linewidth]{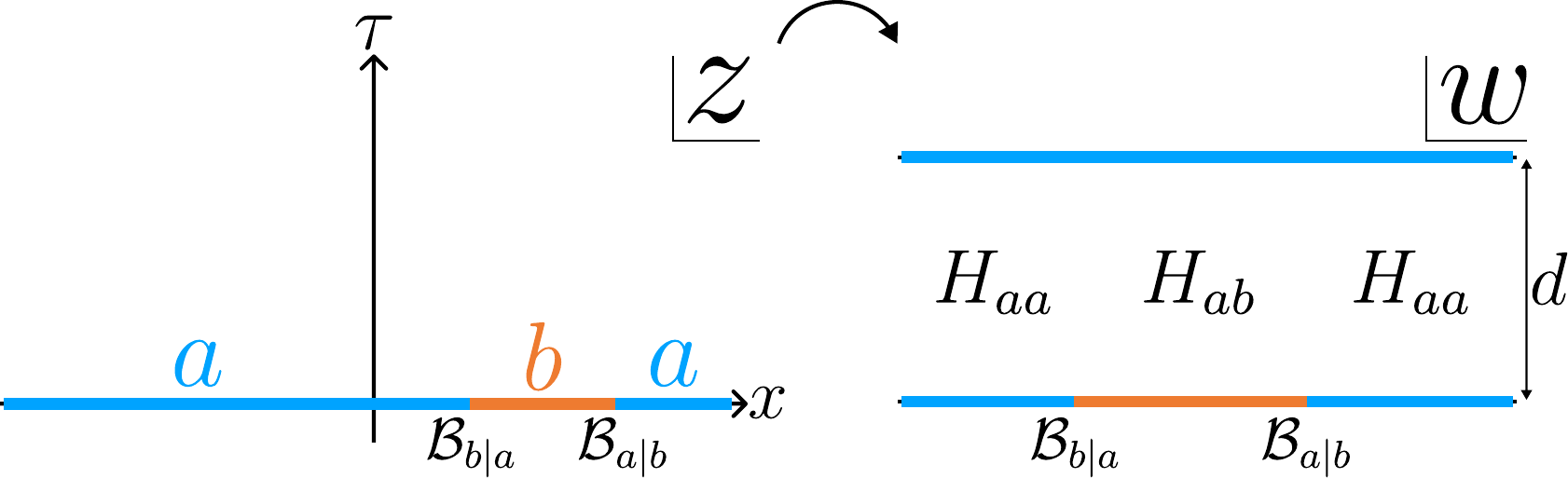}
  \caption{\label{fig:bcco}The upper half-plane maps to an infinite strip under a conformal map $z=de^{\pi w/d}$.
  Quantization along the strip gives two different Hamiltonians $H_{ab}$ and $H_{aa}$ depending on the boundary conditions.}
\end{figure}

In this Appendix, we review the procedure to calculate the scaling dimension of a BCCO~\cite{affleck1997boundary}.
Consider a BCCO $\mathcal{B}_{b\vert a}(x)$ which changes the boundary condition from $a$ to $b$ on the upper half-plane ($\tau\geq0$ with coordinates $z=x+i\tau$).
If the boundary condition $b$ is imposed on the interval $[x_1,x_2]$, then the partition function of this theory is calculated from the two-point correlation function of the BCCOs as:
\begin{equation}\label{eq:bcco-two-point}
  \ev*{\mathcal{B}_{b\vert a}(x_1)\mathcal{B}_{a\vert b}(x_2)} = \frac{1}{(x_1-x_2)^{2\Delta_{b\vert a}}}.
\end{equation}
Under a conformal map $z=de^{\pi w/d}$ that maps the upper half-plane to an infinite strip with finite width $d$, the two-point correlation function [Eq.~\eqref{eq:bcco-two-point}] transforms as:
\begin{align}\label{eq:two-point-1}
      & \ev*{\mathcal{B}_{b\vert a}(u_1)\mathcal{B}_{a\vert b}(u_2)}\notag                                                                             \\
  =   & \qty(\frac{dw}{dz})_{z=x_1}^{-\Delta_{b\vert a}}\qty(\frac{dw}{dz})_{z=x_2}^{-\Delta_{b\vert a}}\frac{1}{(x_1-x_2)^{2\Delta_{b\vert a}}}\notag \\
  =   & \qty[\frac{2d}{\pi}\sinh\frac{\pi}{2d}(u_1-u_2)]^{-2\Delta_{b\vert a
  }}\notag                                                                                                                                             \\
  \to & \qty(\frac{\pi}{d})^{2\Delta_{b\vert a}} \exp(-\frac{\pi\Delta_{b\vert a}}{d}(u_1-u_2))\quad (u_1-u_2\to\infty).
\end{align}
Here, $x_1=de^{\pi u_1/d}$ and $x_2=de^{\pi u_2/d}$.
If we consider the quantization in such a way that the horizontal axis in the strip corresponds to the imaginary time (cf.~Fig.~\ref{fig:bcco}), then the two-point correlation function can be also calculated as follows:
\begin{align}\label{eq:two-point-2}
      & \ev*{\mathcal{B}_{b\vert a}(u_1)\mathcal{B}_{a\vert b}(u_2)}
  = \ev*{\mathcal{B}_{b\vert a}(0)\mathcal{B}_{a\vert b}(\Delta u)}\notag                                                   \\
  =   & \bra*{aa;0}\mathcal{B}_{b\vert a}(0)e^{-H_{ab}\Delta u}\mathcal{B}_{a\vert b}(0)e^{H_{aa}\Delta u}\ket*{aa;0}\notag \\
  =   & \sum_n\abs*{\bra*{aa;0}\mathcal{B}_{b\vert a}(0)\ket*{ab;n}}^2 e^{-[E_{ab}^n(d) - E_{aa}^0(d)]\Delta u}\notag       \\
  \to & \abs*{\bra*{aa;0}\mathcal{B}_{b\vert a}(0)\ket*{ab;0}}^2 e^{-[E_{ab}^0(d) - E_{aa}^0(d)]\Delta u}.
\end{align}
Here, $H_{ab}$ ($H_{aa}$) is the Hamiltonian of the segment of length $d$ with boundary conditions $a$ and $b$ ($a$ and $a$) on each side, $E_{ab}^n(d)$ [$E_{aa}^n(d)$] is the $n$th energy eigenvalue of this Hamiltonian, and $\ket*{ab;n}$ ($\ket*{aa;n}$) is the corresponding eigenstate (see Fig.~\ref{fig:bcco}).
Comparing the two expressions (Eq.~\eqref{eq:two-point-1} and Eq.~\eqref{eq:two-point-2}) for the two-point correlation function, we obtain the following relation:
\begin{equation}\label{eq:bcco-relation}
  \frac{\pi\Delta_{b\vert a}}{d} = E_{ab}^0(d) - E_{aa}^0(d).
\end{equation}
Thus, we can calculate the scaling dimension $\Delta_{b\vert a}$ of a BCCO $\mathcal{B}_{b\vert a}$ from the ground-state energy of the CFT on an infinite strip of width $d$, where the boundary conditions $a$ and $b$ are imposed on each side.

\section{\label{appendix:ee}Entanglement entropy in CFT}

\begin{figure}[tb]
  \includegraphics*[width=0.8\linewidth]{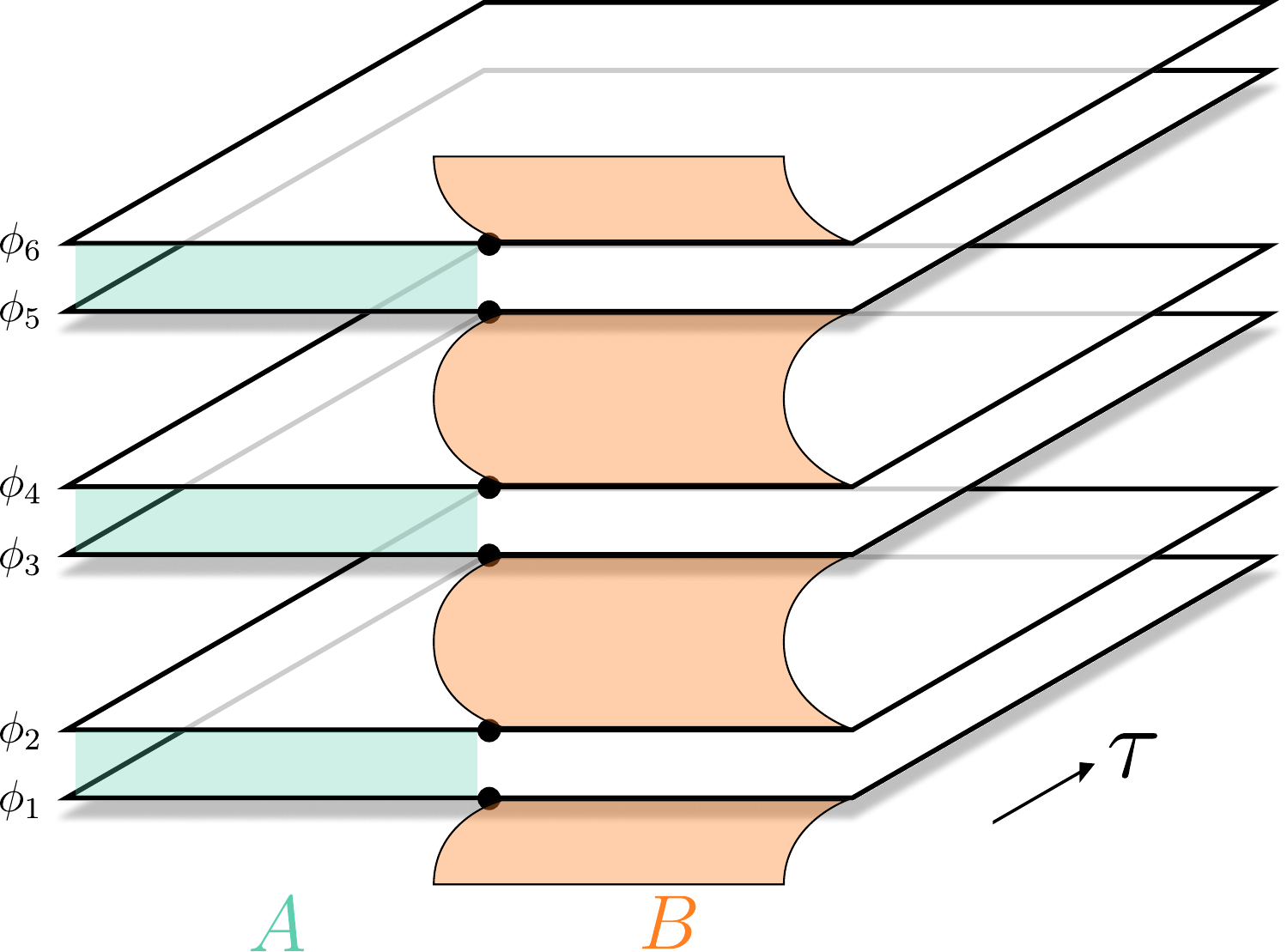}
  \caption{\label{fig:boundary_1}The two boundary conditions $A$ and $B$. $A$ corresponds to taking the partial trace and $B$ is the sewing condition.}
\end{figure}

Here, we derive the well-known result~\cite{HOLZHEY1994443,calabrese2004entanglement} on the entanglement entropy of a finite interval in a CFT, in terms of BCCOs.
Although our derivation is essentially identical to the original ones, it is clarifying and also allows a systematic generalization to a variety of problems as we discussed in the main text.
Let us assume that a CFT is described in terms of a field $\phi$.
The entanglement entropy $S_I=-\Tr[\rho_I\ln \rho_I]$ of the subsystem $I$ can be obtained from the $n{\to}1$ limit of $-\partial_n\Tr[(\rho_I)^n]$.
To calculate $\Tr[(\rho_I)^n]$, we use the replica trick and the folding trick, i.e., we consider the partition function of $2n$ replicas as follows:
\begin{equation}
  \Tr[(\rho_I)^n] = \frac{Z_n(I)}{Z^n}.
\end{equation}
Here $Z_n(I)$ is the path integral of $2n$ replicas $(\phi_1,\phi_2,\ldots,\phi_{2n})$ with the boundary conditions
\begin{align}
  A & :\quad \phi_{2j-1} = \phi_{2j}\; (x\notin I), \\
  B & :\quad \phi_{2j} = \phi_{2j+1}\;(x\in I).
\end{align}
Here, $\phi_{2n+1}\equiv\phi_{1}$.
The partition function $Z_n(I)$ can be interpreted as a two-point function of BCCOs $\mathcal{B}_{A\vert B}$ inserted at the ends of $I$ (see Fig.~\ref{fig:boundary_1}).
The original partition function $Z$ does not depend on the subsystem $I$ in this case, and only serves as a normalization factor.

\begin{figure}[tb]
  \centering
  \includegraphics[width=0.85\linewidth, clip]{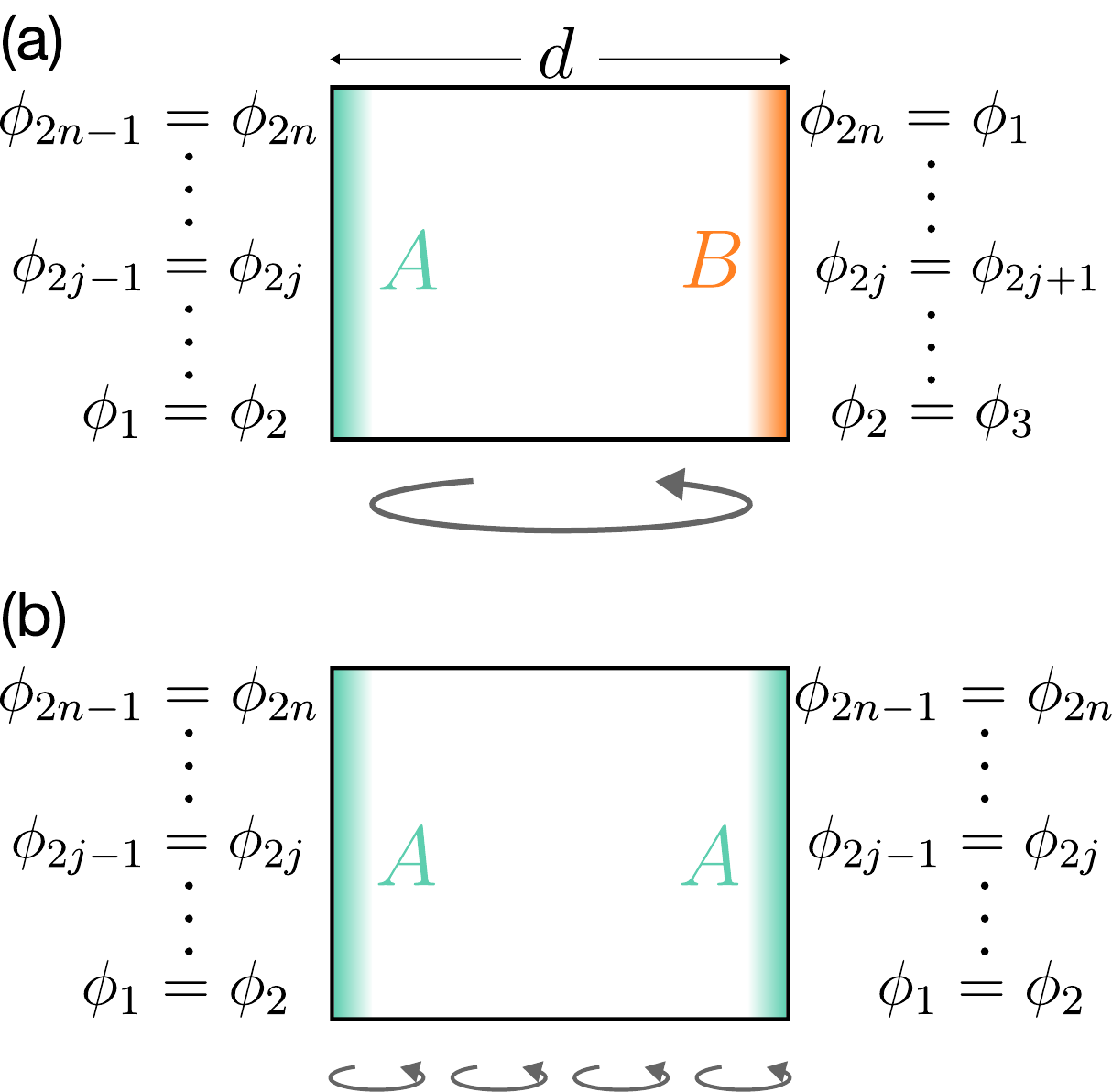}
  \caption{(a)~The strip geometry with boundary condition $A$ on the left and $B$ on the right.
    The $2n$ replicas form a single loop.
    (b)~The strip geometry with boundary condition $A$ on both sides.
    The $2n$ replicas form $n$ loops by pairing up with their partners.}
  \label{fig:gs_energy_ee}
\end{figure}

As explained in the previous section, the scaling dimension of the BCCO $\mathcal{B}_{A\vert B}$ is related to the ground-state energy $E_{AB}^0(d)$ of the CFT on an infinite strip of width $d$, where the boundary conditions $A$ and $B$ are imposed on each side (see Fig.~\ref{fig:gs_energy_ee}).
Suppose that we impose the boundary condition $A$ on the left and $B$ on the right.
Let us start from $\phi_1$ and follow it to the left.
At the left boundary, this is connected to $\phi_2$ as $\phi_1=\phi_2$.
Now following $\phi_2$ to the right, it is connected to $\phi_3$.
Repeating this process $2n$ times, we return to $\phi_1$.
Namely, the replicas form a single loop.
Thus, the ground-state energy $E_{AB}^0(d)$ is identical to the ground-state energy of the original (single-component) CFT on a cylinder of circumference $2dn$:
\begin{equation}
  E_{AB}^0(d) = -\frac{\pi}{6}\frac{c}{2dn} = -\frac{\pi c}{12dn}.
\end{equation}
This is to be compared with the ground-state energy $E_{AA}^0(d)$ of the $2n$-component CFT on the same strip geometry with the boundary condition $A$ on both sides.
Each component of the field is coupled to its partner ($\phi_{2j-1}$ and $\phi_{2j}$) at both sides.
Now, the replicas form $n$ loops by paring up with partners.
Thus, the ground-state energy $E_{AA}^0(d)$ is given by $n$ times the ground-state energy of the original CFT on a cylinder of circumference $2d$:
\begin{equation}
  E_{AA}^0(d) = -\frac{\pi}{6}\frac{c}{2d}\times n = -\frac{\pi nc}{12d}.
\end{equation}

The scaling dimension $\Delta_{A\vert B}$ of the BCCO $\mathcal{B}_{A\vert B}$ is now calculated from Eq.~\eqref{eq:bcco-relation} as
\begin{equation}
  \frac{\pi \Delta_{A|B}}{d} = E_{AB}^0(d) - E_{AA}^0(d) = \frac{\pi c}{12d}\qty(n-\frac{1}{n}),
\end{equation}
which implies
\begin{equation}
  \Delta_{A|B} = \frac{c}{12}\qty(n-\frac{1}{n}).
\end{equation}
Using the BCCOs, the partition function $Z_n(I)$ reads
\begin{equation}
  Z_n(I) = \ev*{\mathcal{B}_{A|B}(l)\mathcal{B}_{B|A}(0)} = l^{-2\Delta_{A|B}} = l^{-\frac{c}{6}\qty(n-\frac{1}{n})},
\end{equation}
leading to the well-known formula for general CFTs:
\begin{equation}
  S_I = \lim_{n\to1}-\frac{\partial}{\partial n}\ln \frac{Z_n(I)}{Z^n}= \frac{c}{3}\ln l + \text{const.}
\end{equation}
Since the original partition function $Z$ is independent of the interval length $l$, the denominator only gives a nonuniversal constant to the entanglement entropy $S_I$.
We note that the BCCO $\mathcal{B}_{B\vert A}$ has been discussed in the context of $S^n$ orbifolds~\cite{lunin2001correlation}, where the scaling dimension was derived in a different way from ours.

\section{Details of numerical calculations}

\begin{figure}[tb]
  \includegraphics*[width=0.95\linewidth,clip]{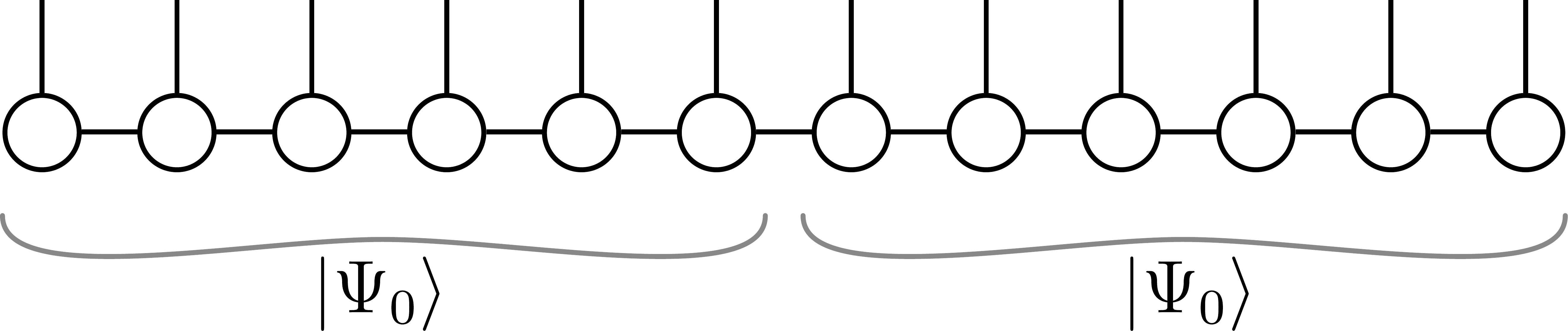}
  \caption{\label{fig:mps_gs}The initial state $\ket*{\Psi_0}\!\otimes\!\ket*{\Psi_0}$ represented as a $2L$-qubit MPS. This MPS is obtained by running the DMRG algorithm with the Hamiltonian $H\otimes I+I\otimes H$.}
\end{figure}

Our numerical calculations use matrix product states (MPS) to represent many-body wave functions.
The measurement process is represented as a contraction of matrix product operators (MPOs) with MPS.
The initial state $\ket*{\Psi_0}\!\otimes\!\ket*{\Psi_0}$ is prepared in $2L$-qubit MPS as shown in Fig.~\ref{fig:mps_gs}.
This state is the ground state of the Hamiltonian $H\otimes I+I\otimes H$, obtained by running the DMRG algorithm.
Here, $H$ is the $L$-qubit Hamiltonian of the critical spin chain.
We can use the same method for other models simply by changing the Hamiltonian.

\begin{figure}[tb]
  \includegraphics*[width=\linewidth,clip]{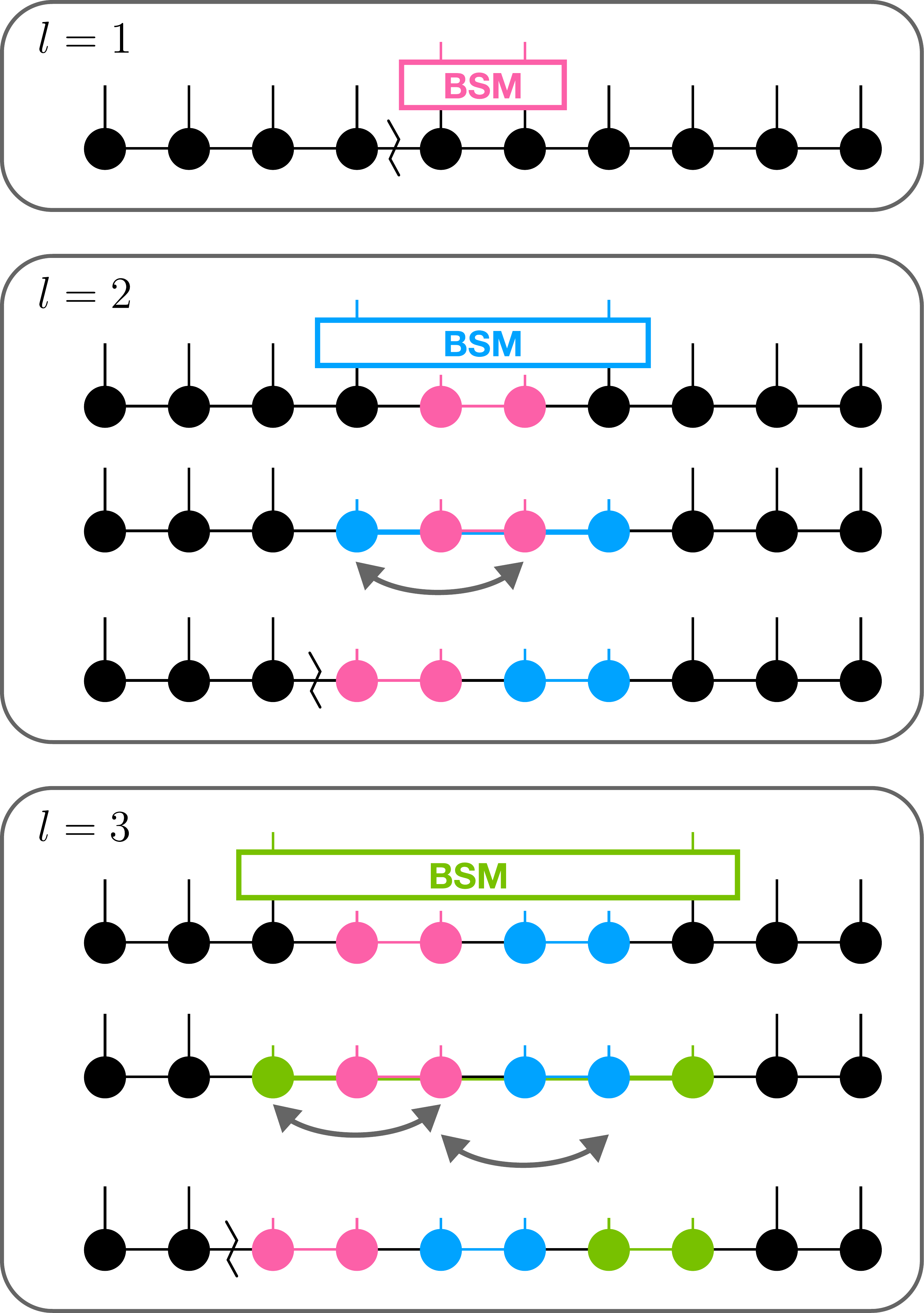}
  \caption{\label{fig:mps_bsm}Steps to avoid the exponentially growing bond dimension in the center. After creating a Bell state with measurement, we switch site labels so that the Bell states do not overlap. The zig-zag displays the bipartitioning point.}
\end{figure}

Measurements are performed by expressing the projection operator $P^{b_1b_2}_{j,j'}=\ketbra*{\mathrm{Bell}^{b_1b_2}_{jj'}}{\mathrm{Bell}^{b_1b_2}_{jj'}}$ with an MPO.
We measure the qubit pairs in the order $(L,L+1),(L-1,L+2),\ldots,(L-l+1,L+l)$, so that the subsystems $A$ and $A'$ become $(1,2,\ldots,L-l)$ and $(L+l+1,\ldots,L)$, respectively.
If we naively continue this procedure, then the bond dimension between qubits $L$ and $L+1$ becomes exponentially large as $l$ increases, making the calculation very inefficient.
This is because the entanglement between both sides is large due to the Bell states created by the measurements.
To avoid this difficulty, we switch the site labels so that the center bond does not carry Bell states on both sides (see Fig.~\ref{fig:mps_bsm}).
Switching site labels in MPS can be done at an affordable cost with singular value decompositions (SVD).

Finally, we calculate the entanglement entropy by bipartitioning the MPS into $A$ and the rest of the chain.
Although this bipartition includes regions $\overline{A},\overline{A'}$, it gives the same EE as $S^{\vec{m}}_A$.
In this way, we can efficiently calculate the swapped entanglement while dealing with two spin chains.

\section{\label{appendix:averaged}Boundary conditions for the averaged swapped entanglement}
As discussed in the main text, the averaged swapped entanglement $\overline{S}_A$ can be calculated by the replica trick, where the measurement-induced boundary interactions are expected to lead to certain conformal boundary conditions.
Here, we discuss the possible boundary condition realized in this replicated theory by focusing on the case of the XXZ critical chains.

To this end, we recall that the averaged swapped entanglement is expressed as follows:
\begin{align}
    & \overline{S}_A\notag                                                                                                                                   \\
  = & \lim_{n\to1}\frac{1}{1-n}\sum_{\vec{m}}p_{\vec{m}}\ln\Tr[(\rho^{\vec{m}}_A)^n]\notag                                                                   \\
  = & \lim_{n\to1}\frac{1}{1-n}\sum_{\vec{m}}\Tr[\tilde{\rho}^{\vec{m}}]\qty(\ln\Tr[(\tilde{\rho}_A^{\vec{m}})^n]- \ln(\Tr[\tilde{\rho}^{\vec{m}}])^n)\notag \\
  = & \lim_{\substack{n\to1                                                                                                                                  \\s\to0}}\frac{1}{s(1-n)}\sum_{\vec{m}}\Tr[\tilde{\rho}^{\vec{m}}]\qty(\Tr[(\tilde{\rho}_A^{\vec{m}})^n]^s - \Tr[\tilde{\rho}^{\vec{m}}]^{ns})\notag \\
  = & \lim_{\substack{n\to1                                                                                                                                  \\s\to0}}\frac{1}{s(1-n)}\sum_{\vec{m}}\qty[\Tr[\tilde{\rho}^{\vec{m}}]\Tr[(\tilde{\rho}_A^{\vec{m}})^n]^s - \Tr[\tilde{\rho}^{\vec{m}}]^{ns+1}].
\end{align}
The boundary action $\delta\mathcal{S}$ for the $R=ns{+}1$ replicated theory can be extracted from this expression as
\begin{align}
  e^{-\delta\mathcal{S}}
   & = \sum_{\vec{m}}(P^{\vec{m}})^{\otimes R} = \sum_{\vec{m}}(P^{m_1})^{\otimes R}\cdots(P^{m_l})^{\otimes R}\notag \\
  \delta\mathcal{S}
   & = -\sum_{j=1}^l\ln\sum_{m_j}(P^{m_j})^{\otimes R}.
\end{align}
We want to rewrite the sum of four operators $\sum_m(P^{m})^{\otimes R}$ into a single exponential.
This can be done by expressing the operator $(P^{00})^{\otimes R}$ as a sum of the elements in a stabilizer group $\ev*{S}$, where the generator $S$ is
\begin{equation}
  S = \{XX^{(1)},ZZ^{(1)},\ldots,XX^{(R)},ZZ^{(R)}\}.
\end{equation}
Here, $XX^{(k)},ZZ^{(k)}$ is the product of two Pauli matrices in the $k$th replica.
The other three operators ($m=01,10,11$) have the same stabilizer group, but with different signs for $XX$s and $ZZ$s.
Thus, if we take the summation over the measurement outcomes, then only the elements generated with an even number of $XX$s and $ZZ$s each can remain.
The sum of the operators can then be expressed as a sum of the elements in a stabilizer group $\ev*{\Tilde{S}}$, where the generator $\tilde{S}$ is
\begin{equation}
  \tilde{S} = \{XX^{(1)}XX^{(2)}, ZZ^{(1)}ZZ^{(2)},\ldots,ZZ^{(R-1)}ZZ^{(R)}\}.
\end{equation}
From this expression, we can rewrite the sum of the operators in a single exponential:
\begin{align}
  \sum_m (P^{m})^{\otimes R}
   & = \frac{1}{2^{2R-2}}\sum_{M\in\ev*{\tilde{S}}}M\notag \\
   & = \prod_{s\in\tilde{S}}\frac{1+s}{2}\notag            \\
   & \propto\lim_{g\to\infty}\exp(g\sum_{s\in\tilde{S}}s)
\end{align}
This results in the following boundary action:
\begin{align}
  \delta S = -\lim_{g\to\infty}g & \int_0^l\!dx\int\!d\tau  \delta(\tau) \; (XX^{(1)}XX^{(2)}\notag \\
  +                              & ZZ^{(1)}ZZ^{(2)} + \cdots + ZZ^{(R-1)}ZZ^{(R)}).
\end{align}
To ensure the $U(1)$ symmetry, we can safely add the $YY^{(k)}YY^{(k+1)}$ term as adding this term does not alter the resulting operator in the limit $g\to\infty$.
Using the bosonization relation:
\begin{align}
  XX^{(k)} & \sim \cos\theta_-^{(k)} + \cos\theta_+^{(k)}, \\
  YY^{(k)} & \sim \cos\theta_-^{(k)} - \cos\theta_+^{(k)}, \\
  ZZ^{(k)} & \sim \cos 2\phi_+^{(k)} + \cos 2\phi_-^{(k)},
\end{align}
we can write the boundary action in terms of the boson fields $\phi_{\pm}^{(k)},\theta_{\pm}^{(k)}$ as
\begin{widetext}
  \begin{align}
            & XX^{(1)}XX^{(2)} + YY^{(1)}YY^{(2)}\notag                                                                                                                                                              \\
    \propto & \cos\theta_-^{(1)}\cos\theta_-^{(2)} + \cos\theta_+^{(1)}\cos\theta_+^{(2)}\notag                                                                                                                      \\
    \propto & \cos \theta_-^{(1)}{+}\theta_-^{(2)} + \cos \theta_-^{(1)}{-}\theta_-^{(2)} + \cos \theta_+^{(1)}{+}\theta_+^{(2)} + \cos \theta_+^{(1)}{-}\theta_+^{(2)}\notag                                        \\
    =       & \cos(\theta_1{-}\theta_2{+}\theta_3{-}\theta_4) + \cos(\theta_1{-}\theta_2{-}\theta_3{+}\theta_4) + \cos(\theta_1{+}\theta_2{+}\theta_3{+}\theta_4) + \cos(\theta_1{+}\theta_2{-}\theta_3{-}\theta_4), \\
    \text{and}\notag                                                                                                                                                                                                 \\
            & ZZ^{(1)}ZZ^{(2)}\notag                                                                                                                                                                                 \\
    \propto & (\cos 2\phi_+^{(1)} + \cos 2\phi_-^{(1)})(\cos 2\phi_+^{(2)} + \cos 2\phi_-^{(2)})\notag                                                                                                               \\
    \propto & \cos 2(\phi_1{+}\phi_2{+}\phi_3{+}\phi_4)+\cos 2(\phi_1{+}\phi_2{-}\phi_3{-}\phi_4)+\cos 2(\phi_1{-}\phi_2{+}\phi_3{+}\phi_4)+\cos 2(\phi_1{-}\phi_2{-}\phi_3{-}\phi_4)\notag                          \\
            & +\cos 2(\phi_1{+}\phi_2{+}\phi_3{-}\phi_4)+\cos 2(\phi_1{+}\phi_2{-}\phi_3{+}\phi_4)+\cos 2(\phi_1{-}\phi_2{+}\phi_3{-}\phi_4)+\cos 2(\phi_1{-}\phi_2{-}\phi_3{+}\phi_4).
  \end{align}
\end{widetext}
Here, we wrote $\theta_1^{(1)}=\theta_1,\theta_2^{(1)}=\theta_2,\theta_1^{(2)}=\theta_3,\theta_2^{(2)}=\theta_4$ and the same for $\phi$.

From the scaling dimensions ($2K$ for $\phi$ and $1/2K$ for $\theta$), we may argue that the condition on $\theta$ is the most relevant one.
Thus, the resulting boundary condition should be $\theta_1^{(k)}=\theta_2^{(k)}=0$ for all the replicas.
This is the Neumann boundary condition for $\phi$, which is conformally invariant.
Interestingly, all the $\phi$ components end up completely decoupled through the interactions induced by the measurements.
To test the above prediction, we need to determine the scaling dimensions of the corresponding BCCOs, which we leave to a future work.

\section{\label{appendix:general}Entanglement swapping in general CFTs}

\begin{figure*}[tb]
  \includegraphics*[width=\linewidth,clip]{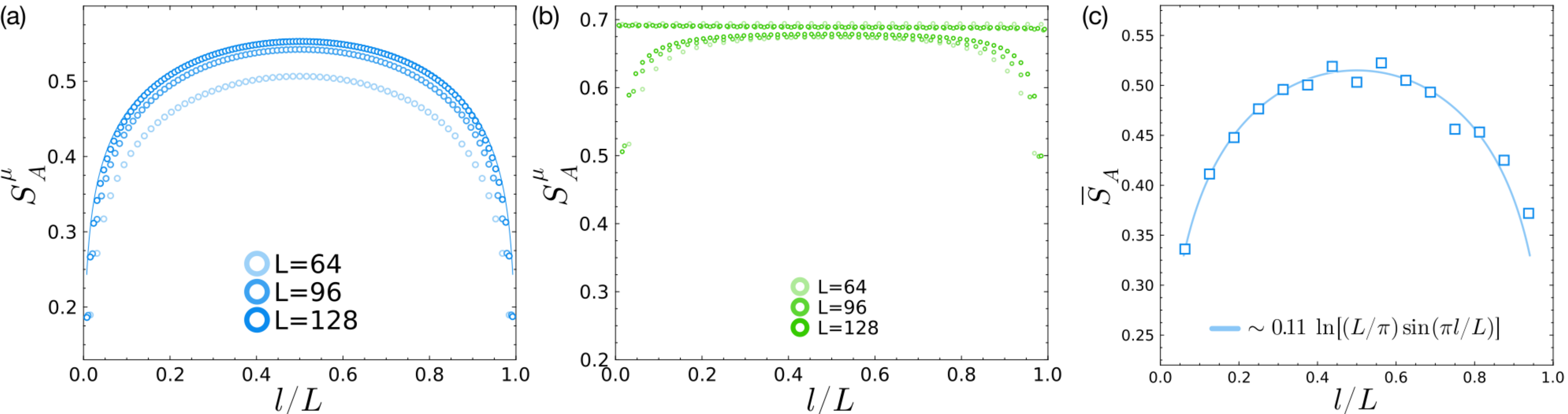}
  \caption{\label{fig:tfim}
  (a)~Numerical results of $S^{\vec{\mu}}_A$ for $\mu=00$ in the TFIM.
  Data are fitted to the curve $S^{\vec{\mu}}_A=c_{\vec{\mu}}\ln[(L/\pi)\sin(\pi l/L)]+\text{const}$, leading to $c_{\vec{\mu}}=0.097$.
  The constant is the same for all four curves.
  (b)~Numerical results of $S^{\vec{\mu}}_A$ for $\mu=01$.
  (c)~Numerical results of $\overline{S}_A$. Statistical errors are small compared to the size of the markers.}
\end{figure*}

\subsection{Time-reversal symmetric states}
We can exactly calculate the swapped EE in the case of uniform measurement outcomes of $\mu=00$, assuming the time-reversal symmetry of the initial state.
Let us use the Schmidt decomposition and write the initial state $\ket*{\Psi_0}$ in the following form:
\begin{equation}
  \ket*{\Psi_0} = \sum_{i=1}^{\chi} \sqrt{p_i}\ket*{\overline{\alpha}_i}\ket*{\alpha_i}.
\end{equation}
Here, $\sqrt{p_i}$ is the Schmidt coefficient, $\chi$ is the Schmidt rank, and $\ket*{\alpha_i}$ ($\ket*{\overline{\alpha}_i}$) represent the states in the subsystem $A$ ($\overline{A}$).
From this expression of the initial state, we can write the postmeasurement state $\ket*{\Psi_{AA'}^{\vec{\mu}}}$ as follows
\begin{align}
  \sqrt{p_{\vec{\mu}}}\ket*{\Psi_{AA'}^{\vec{\mu}}}
   & = \bra*{\mathrm{Bell}^{00}}^{\otimes l}\ket*{\Psi_0}\!\ket*{\Psi_0}\notag                                                                                                                                     \\
   & = \sum_{i,j=1}^{\chi}\sqrt{p_ip_j} \underbrace{\qty(\bra*{\mathrm{Bell}^{00}}^{\otimes l}\ket*{\overline{\alpha}_i}\!\ket*{\overline{\alpha}_j})}_{2^{-l/2}\delta_{ij}}\ket*{\alpha_i}\!\ket*{\alpha_j}\notag \\
   & = 2^{-l/2}\sum_{i=1}^{\chi}p_i \ket*{\alpha_i}\!\ket*{\alpha_i}.
\end{align}
In the second line, we used the relation $\braket*{\mathrm{Bell}^{00}}{\psi}\!\ket*{\psi}=2^{-1/2}\braket*{\psi^\ast}{\psi}$, and the time-reversal symmetry $\ket*{\overline{\alpha}_i^\ast}=\ket*{\overline{\alpha}_i}$.
Thus, the Born probability $p_{\vec{\mu}}$ can be written as $p_{\vec{\mu}}=2^{-l}\sum_i p_i^2 = e^{-(\ln 2)l} e^{-S_A(2)}$, where $S_A(n)=(1-n)^{-1}\ln \sum_i p_i^n$ is the R\'{e}nyi entropy of the subsystem $A$ in the initial state.
For the ground state of a CFT, it is known to behave as $S_A(n)=(c/6)(1+1/n)\ln l$ in the long-distance limit.
Furthermore, the swapped EE can be calculated from the following expression of the reduced density matrix:
\begin{equation}
  \rho_A^{\vec{\mu}} = \Tr_{A'}[\ketbra*{\Psi_{AA'}^{\vec{\mu}}}{\Psi_{AA'}^{\vec{\mu}}}] = \sum_i\frac{p_i^2}{\sum_j p_j^2}\ketbra*{\alpha_i}{\alpha_i},
\end{equation}
which is diagonal.
Therefore,
\begin{align}
  S_A^{\vec{\mu}}(n)
   & := \frac{1}{1-n}\ln\Tr[(\rho_A^{\vec{\mu}})^n]\notag                     \\
   & = \frac{1}{1-n}\ln \frac{\sum_i p_i^{2n}}{(\sum_j p_j^2)^n}\notag        \\
   & = \frac{1}{1-n}\ln\sum_i p_i^{2n} - \frac{n}{1-n}\ln \sum_i p_i^2 \notag \\
   & = \frac{1-2n}{1-n} S_A(2n) + \frac{n}{1-n} S_A(2),
\end{align}
and in the long-distance limit,
\begin{equation}
  S_A^{\vec{\mu}}(n) \sim \frac{c}{12}\qty(1 + \frac{1}{n})\ln l.
\end{equation}
By taking the $n\to1$ limit, we obtain the swapped EE for general CFTs:
\begin{equation}
  S_A^{\vec{\mu}} \sim \frac{c}{6}\ln l.
\end{equation}
The result is consistent with the case of the XXZ chain ($c=1$ free-boson CFT).

The coefficient $c/6$ of the swapped EE for general CFTs can also be derived by using field-theoretical arguments.
We assume that the boundary condition imposed by an uniform measurement in $\ket*{\mathrm{Bell}^{00}}=(\ket*{00}+\ket*{11})/\sqrt{2}$ leads to the boundary condition $\phi=\phi'$ for any CFTs.
This assumption can be understood intuitively, as this measurement forces the two qubits to occupy the same state (either $\ket*{0}$ or $\ket*{1}$) without constraining any specific spin configuration.
Under this assumption, the calculations for the free-boson CFT in the main text can be extended to any CFTs by changing the central charge from $c=1$ to a general value $c$.
This is possible because the Casimir energy only depends on the central charge, and does not rely on further information of the specific CFT we consider.

\subsection{Ising CFT}
We can check the result in the previous subsection by considering another prototypical example of a critical spin chain; the critical transverse field Ising model (TFIM), which realizes a $c=1/2$ Ising CFT.
The lattice Hamiltonian of the critical TFIM is given as
\begin{equation}
  H = -\sum_{j=1}^{L}(\sigma_j^z\sigma_{j+1}^z + \sigma_j^x).
\end{equation}
The ground state of the TFIM respects the $\mathbb{Z}_2$ symmetry, i.e., $G\ket*{\Psi_0}=\ket*{\Psi_0}$ where $G=\prod_{j}\sigma_j^x$.
Thus, the two pairs of the Bell bases $ \mu=00,10$ and $\mu=01,11$ give the same swapped entanglement $S^{\vec{\mu}}_A$ because they can be related to each other by the unitary $I\otimes G$.

Numerical calculations of $S^{\vec{\mu}}_A$ for $\mu=00$ and $\mu=01$ are shown in Figs.~\ref{fig:tfim}(a) and \ref{fig:tfim}(b).
For $\mu=00$, the swapped entanglement scales logarithmically with $l$, implying that the boundary condition $\vec{\mu}$ for $\mu=00$ is conformally invariant.
The extracted coefficient $c_{\vec{\mu}}=0.097$ is not too far from the theoretical value $c_{\vec{\mu}}=1/12$.
Meanwhile, for $\mu=01$, we find a qualitatively different behavior.
Namely, when the number of measured qubits $l$ is odd, the swapped entanglement precisely takes a value of $S^{\vec{\mu}}_A=\ln2$, while it saturates towards a constant for large $L$ when $l$ is even.
This implies that boundary conditions induced by the measurement outcomes $\mu=01$ and $\mu=11$ are not likely to be conformally invariant.

We have also calculated the averaged swapped EE $\overline{S}_A$ as shown in Fig.~\ref{fig:tfim}(c).
Despite the fact that the boundary conditions induced by the measurement outcomes $\mu=01$ and $\mu=11$ should not be conformally invariant, the averaged value $\overline{S}_A$ recovers the logarithmic scaling with $l$.
Interestingly, the value of the coefficient in the logarithmic scaling is close to that of the uniform outcome case, in the same manner as in the case of the TLLs discussed in the main text.
Theoretically explaining this interesting behavior remains an open question.

%%% bibliography %%%


\end{document}